\documentclass[prd,aps,floatfix,nofootinbib,preprint,tightenlines,superscriptaddress]{revtex4-2}
\usepackage{dcolumn}
\usepackage{amsmath}
\usepackage{latexsym}
\usepackage{graphicx}
\usepackage{bm}

\def\co{{\cal O}}
\def\bx{{\bf x}}

\def\svev#1{\left\langle #1\right\rangle}       

\def\Tr{{\rm Tr}\,}

\newcommand{\bee}{\begin{equation}}
\newcommand{\ee}{\end{equation}}
\newcommand{\beea}{\begin{eqnarray}}
\newcommand{\eea}{\end{eqnarray}}

\begin{document}

\title{Lattice study of the chiral  properties of large $N_c$ QCD}

\author{Thomas DeGrand, Evan Wickenden}
\affiliation{
Department of Physics, University of Colorado,
        Boulder, CO 80309 USA}
\email{thomas.degrand@colorado.edu}

\date{\today}

\begin{abstract}
We present a lattice calculation of the low energy constants  of 
QCD with $N_c=3$, 4 and 5 colors and $N_f=2$ flavors of degenerate mass fermions.
We fit data for the pseudoscalar meson mass, the pseudoscalar decay constant, and the Axial Ward Identity
fermion mass to formulas from next to next to leading order chiral perturbation theory.
We extract the next to leading order low energy constants and study their behavior as a function of $N_c$.
Pre-existing analyses of $N_c=3$ inform our fitting strategies.
\end{abstract}

\maketitle


\section{Introduction and motivation}

The most studied deformation of real-world three-color ($N_c=3$) QCD is its extension to a large number of colors.
In the limit that $N_c$ is taken to infinity, it is believed that mesons are narrow quark anti-quark
bound states, baryon masses scale as $N_c$, and  $N_c$ counting rules set the overall scale for hadronic matrix elements
\cite{tHooft:1973alw,tHooft:1974pnl,Witten:1979kh}.

There is a small lattice literature related to the large $N_c$ limit of QCD. 
Simulations involve
the pure gauge theory and fermions included in quenched approximation
 (mostly at single-digit values of $N_c$), 
quenched simulations at very large $N_c$ and small volume, and simulations in large volume
with a small number of dynamical fermion flavors (so far,
$N_f=2$ and 4).
 Refs.~\cite{Lucini:2012gg,GarciaPerez:2020gnf,Hernandez:2020tbc}
are a selection of reviews of the various approaches. The goal of these simulations is to confront predictions of
large $N_c$ QCD, which are typically (semi) analytic, with nonperturbative lattice-based results.

The qualitative situation with regard to such comparisons is as follows:
Choose to fix the lattice spacing in a simulation using some observable 
taken from a correlation function dominated by gluonic degrees of freedom.
 Such observables include the string tension, 
the Sommer parameter \cite{Sommer:1993ce},  or the flow \cite{other,Luscher:2010iy}
parameter $t_0$.
Perform lattice simulations while tuning the bare parameters (the most sensitive
 one is the gauge coupling $g^2$) so that the lattice spacing
is the same for all  values of $N_c$. One will discover that the bare
 `t Hooft couplings $\lambda=g^2 N_c$  are
roughly matched across $N_c$. Alternatively, performing simulations across $N_c$ at the same value
of $\lambda$, one will discover that  the lattice spacing had been approximately matched.
Once that is done one will see that the masses of flavor nonsinglet mesons,
and their dependence on the fermion mass, will be approximately equal across $N_c$.
Simple matrix elements (decay constants, the kaon $B-$ parameter) will scale with $N_c$ as predicted from the color weight of
amplitudes.
All of these statements are modified by small corrections, which can be arguably interpreted 
as $1/N_c$ effects.

We are not aware of any large volume simulations which attempt to take a continuum limit;
most of them involve simulations across $N_c$ at a single matched value lattice spacing.
It seemed to us that it might be possible to extend one of these studies to do that.
We make a first attempt to  measure at chiral parameters of $SU(N_c)$ gauge theory with
$N_f=2$ flavors of degenerate fermions, in the continuum and large $N_c$ limits.
Specifically, we studied $N_c=3$, 4 and 5. We wanted to address the following set of questions:
\begin{enumerate}
\item How do the low energy constants of the chiral effective theory scale with $N_c$?
\item What is their $N_c\rightarrow \infty$ limit? How do our results compare to those from other
approaches (quenched simulations, simulations in small volume)?
\item As $N_c$ rises, the low energy effective theory is expected to change from the conventional
$SU(N_f)_L\times SU(N_f)_R \rightarrow SU(N_f)_{L+R}$ pattern of symmetry breaking to one where the
flavor singlet meson becomes light
 \cite{DiVecchia:1980yfw,Rosenzweig:1979ay,Witten:1980sp,Kawarabayashi:1980dp,Herrera-Siklody:1996tqr,Kaiser:2000gs}.
At what value of $N_c$ does this ``$U(N_f)$'' expansion begin to reproduce the data?
\end{enumerate}

There is already an extensive literature about the low energy constants of $N_c=3$, $N_f=2$ QCD,
summarized in the Flavor Lattice Averaging Group's 2019 summary,
 Ref.~\cite{FlavourLatticeAveragingGroup:2019iem}.
Typical analyses in their summary have much better quality than what 
we present here; we cannot add much to this already-closed
subject. We instead use the results of these pre-existing  $N_c=3$, $N_f=2$ to inform
our analysis, and to calibrate the accuracy we claim for $N_c> 3$.

We should say at the start that while we have made a first attempt to extract continuum
to the low energy constants of $N_c>3$ QCD, we do not regard our results as definitive.
The main issues are three: two typical of early QCD studies of any observable, with
too large a lattice spacing and too large fermion masses for comfortable extrapolations,
the third is that we do not feel that we have enough values of $N_c$. But the outline of the 
solution is there: once again, $N_c=3$ is not that different from infinite $N_c$.

Our calculation was strongly influenced by two previous ones. The first is a 2013 computation of the 
low energy parameters of $N_f=2$ QCD ($N_c=3$) in the continuum limit, by the 
Budapest-Marseille-Wuppertal collaboration \cite{Budapest-Marseille-Wuppertal:2013vij}.
Their study has larger volumes, smaller lattice spacings, and smaller fermion masses, and
than our exploratory project. We adapted much of their analysis methodology.
The other project was a calculation of the low energy parameters of $N_f=4$ systems
by P.~Hern\'andez, C.~Pena and F.~Romero-L\'opez \cite{Hernandez:2019qed}.
This was  done at a single lattice spacing, matched across $N_c$. They studied
$N_c=3$, 4, 5, and 6. The addition of $N_c=6$ allowed for more controlled extrapolation to
infinite $N_c$ than we could do.
This is because large $N_c$ predictions are typically involve a power series in 
$N_c$:
\bee
\svev{O(N_c)} = c_0 + \frac{c_1}{N_c} + \frac{c_2}{N_c^2}  + \dots
\ee
and it is useful to have more than $j+1$ values of $N_c$ to fit data to such predictions
at $j$th order.

Most discussions of the literature of large $N_c$ QCD (such as the one we just gave) make the point that
 comparisons across $N_c$ reveal only
 small differences. But
to lattice practitioners QCD's at different $N_c$'s are different.
Some of these differences are unfavorable.
For example, the up-front cost of a simulation at some $N_c$ scales roughly like $N_c^2$
 (when matrix times vector multiplication dominates, a situation encountered in inverting
the Dirac operator) or $N_c^3$ (for matrix multiplication of gauge links).
 The simulation autocorrelation time 
as measured by the topological susceptibility grows with $N_c$, in both quenched simulations and ones
with dynamical fermions.

However, differences across $N_c$ can be favorable for larger $N_c$. That is the case in this project,
where we had much greater difficulty generating useful $N_c=3$ data sets than we did with $N_c=4$ or 5.
This is due to the scaling of the pseudoscalar decay constant $F$ with $N_c$: $F\propto \sqrt{N_c}$.
This is helpful in two (related) ways.
First, the expansion parameter of chiral perturbation theory is (speaking loosely) the ratio of
the squared pseudoscalar mass to $F^2$, 
$x=M_{PS}^2/(8\pi^2 F^2)$.  Simulations which make comparisons at identical
 $x$ across $N_c$ involve ever larger values of $M_{PS}^2$ and become easier to perform as $N_c$ grows.
Second, finite volume corrections in a box of size $L$
  are proportional to $\Delta(m,L)/F^2$ where $\Delta(m,L)$ is the
propagator of a pseudoscalar``to an image point,''
which appears along with the  ``tadpole'' or ```snail'' graph in finite volume.
  The $1/F^2\propto 1/N_c$ prefactor means that as $N_c$ grows there are smaller
 finite volume corrections
 at identical pion masses. As a consequence, at bigger $N_c$
one can simulate at smaller volume without encountering volume artifacts.

The outline of the paper is as follows: Sec.~\ref{sec:theory} sets the conventions we
follow for
the low energy chiral effective theory, to which we compare our data.
Secs.~\ref{sec:latticetech} and \ref{sec:fitting} describes the various parts of the lattice simulation,
first the collection of data and its conversion to the dimensionless quantities which
are what we fit, and then a discussion of issues involved in the fits themselves.
Results of fits which extract the low energy constants of the chiral effective Lagrangian
are presented in Sec.~\ref{sec:results}.
Some conclusions are found in Sec.~\ref{sec:conclude}.

\section{Theoretical background and conventions for chiral observables\label{sec:theory}}

The goal of this project is to compare the relationship between quantities
 relevant to the low energy properties of $SU(N_c)$ gauge theory with two flavors of degenerate mass
fundamental representation fermions, across $N_c$ and  in the continuum limit. We focus on 
the fermion mass, the (squared)
pseudoscalar mass  and pseudoscalar decay constant as measured in simulations.
We label these quantities as $m_q$, $m_{PS}^2$, and $f_{PS}$. The comparisons are done
 in the context of the parameters of a low energy effective chiral Lagrangian.
There are two choices for an effective theory, which differ in the number of light degrees of freedom they
represent.

The effective field theory for $N_c=3$ QCD is based on the spontaneous breaking of
chiral symmetry $SU(N_f)_L\times SU(N_f)_R \rightarrow SU(N_f)_{L+R}$ (plus its explicit brealing
by quark masses). Its small expansion parameter is
\bee
O(\delta) \sim O(p^2) \sim O(M^2) \sim O(m_q)
\label{eq:deltasu}
\ee
where $M^2$ is the squared mass of the fields in the Lagrangian (taken to be massless
 Goldstone bosons in the $m_q\rightarrow 0$ limit). We will refer to this effective theory
as ``$SU(N_f)$ effective theory.'' (Of course, for us, $N_f=2$.)

Only flavor nonsinglet mesons appear in the Lagrangian of  $SU(2)$ effective theory. In particular,
the flavor singlet meson, the eta-prime (in $N_f=3$ language) gets a large mass from the anomaly
and does not appear in the low energy theory of the $N_c=3$ world.

However, the situation at large $N_c$ is different. The size of anomaly term falls
 with $N_c$ and the squared eta prime mass decreases as $1/N_c$, so at some point the eta prime must
be included in the low energy effective theory. The resulting theory is called
$U(N_f)$ (here $U(2)$) effective theory and has a slightly different chiral expansion.
Which chiral expansion scheme works better for any particular value of $N_c$ is an open question, which we hope
to explore.

We begin by discussing $SU(2)$ chiral perturbation theory.
We follow the usual convention for lattice simulations and define the  pseudoscalar decay constant $f_{PS}$ in terms
of the matrix element
\bee
\langle 0| \bar u \gamma_0 \gamma_5 d |\pi\rangle = m_{PS} f_{PS}.
\label{eq:fpi}
\ee
This leads to the identification $f_{\pi} \sim 132$ MeV in QCD. Note that the continuum chiral PT literature
uses a ``93 MeV'' definition: the extra $\sqrt{2}$ is an isospin raising factor. See Ref.~\cite{DeGrand:2019vbx}
for a compilation of conventions for this quantity.

There are two
commonly used expansion schemes for $SU(N_f)$ chiral perturbation theory. 
The two leading order constants are $B$ and $F$.
We will mostly use the  ``$x$-expansion'' which is given 
in terms of $m_q$ via the quantity
of
\bee
x= \frac{M^2}{8\pi^2 F^2},
\label{eq:xxxx}
\ee
where $M^2=2B m_q = 2\Sigma m_q/F^2$ and $\Sigma$ is the fermion condensate.
Note that $x$ is $O(\delta)$ according to Eq.~\ref{eq:deltasu}.
 In  next to leading order (NLO), Bijnens and Lu \cite{Bijnens:2009qm}
define (converting $F$ conventions)
\bee
\overline A(M^2) = \frac{M^2}{8\pi^2} \log \frac{\mu^2}{M^2}
\ee
and the formulas to be addressed are
\beea
m_{PS}^2 &=& 2Bm_q \left[ 1 + \frac{a_M}{F^2} \overline A(M^2) + \frac{M^2}{F^2} b_M + \dots \right] \nonumber \\
f_{PS} &=& F\left[ 1 + \frac{a_F}{F^2} \overline A(M^2) + \frac{M^2}{F^2} b_F  + \dots \right] . \nonumber \\
\eea
Here
\beea
a_M &=& -\frac{1}{N_f}  \nonumber \\
b_M &=& 8N_f(2L_6-L_4)+8(2L_8-L_5)  \nonumber \\
a_F &=& \frac{1}{2}N_f  \nonumber \\
b_F &=& 4(N_f L_4+L_5) .  \nonumber \\
\eea
The $L_i$'s are the low energy constants (LEC's) of the NLO chiral expansion.
The $b_j$'s depend on the choice of scale in the logarithm of $\bar A(M^2)$.

The FLAG review  \cite{FlavourLatticeAveragingGroup:2019iem} and BMW
\cite{Budapest-Marseille-Wuppertal:2013vij} make a slightly different
definition, which we will employ:
\bee
 \frac{a_j}{F^2} \overline A(M^2) +
 \frac{M^2}{F^2} b_j \equiv a_j \frac{M^2}{8\pi^2 F^2}\left[  \log \frac {\mu^2}{M^2} + l_j \right].
\label{eq:switch}
\ee
The $l_i$'s are scheme ($\mu^2$) dependent, while $B$ and $F$ are not. Note that
\bee
l_i(\mu_2)=l_i(\mu_1) + \log \frac{\mu_1^2}{\mu_2^2}  .
\label{eq:sushift}
\ee
The convention
 is that $b_M$ is replaced by $l_3$
and $b_F$ is replaced by $l_4$.

We will actually need the next-to next-to leading order (NNLO) expression:
\beea
m_{PS}^2 &=& 2Bm\left[ 1 - \frac{1}{2}x(\ln\frac{\mu^2}{M^2} + l_3) + x^2(\frac{17}{8}T_M^2 + k_M) \right] \nonumber \\
f_{PS} &=& F\left[ 1 + x(\ln\frac{\mu^2}{M^2} + l_4) + x^2(-\frac{5}{4}T_F^2 + k_F) \right]  , \nonumber \\
\label{eq:xfits}
\eea
where
\beea
T_M &=& \ln\frac{\mu_\pi^2}{M^2} + \frac{60}{51}l_{12} - \frac{9}{51}l_3 + \frac{49}{51} \nonumber \\
T_F &=& \ln\frac{\mu_\pi^2}{M^2} + l_{12} + \frac{1}{5}(l_3-l_4) + \frac{23}{30}. \nonumber \\
\label{eq:tftm}
\eea

Our analysis convention is to fix $a_M= -1/2$ and $a_F=1$ in any fit. Then
at NLO there are four LEC's which can be measured, $B$, $F$, $l_3$ and $l_4$.
NNLO adds an additional three LEC's, ($l_{12}$, $k_M$, $k_F$) to the collection of quantities to be fit.
The LEC's are expected to show the following $N_c$ scaling (the
 quick list can be found in in Ref.~\cite{Hernandez:2019qed}):
\bee
O(N_c): F^2, L_5, L_8, \qquad O(1): B, L_4, L_6 .
\label{eq:scalenc}
\ee
This means that the  $l_i$'s for $i=3,4$ scale as
\bee
l_i= N_c l_i^{(0)} + l_i^{(1)} .
\label{eq:linc}
\ee
Our analysis of data using $SU(N_f)$ chiral perturbation theory is done with separate fits for each $N_c$.
We can then ask whether the LEC's scale with $N_c$ as given by Eqs.~\ref{eq:scalenc}-\ref{eq:linc}.

The ``$\xi$-convention'' rewrites the chiral expansion in terms of a ratio of observables
\bee
\xi = \frac{m_{PS}^2}{8\pi^2 f_{PS}^2}
\label{eq:xi}
\ee
(in the ``132 MeV'' definition for the decay constant). We have only implemented the NLO expressions,
which are
\bee
M^2 = 2Bm_q= m_{PS}^2\left[ 1 + \frac{1}{2}\xi (\ln\frac{\mu^2}{m_{PS}^2} + l_3) \dots \right]
\label{eq:bxifit}
\ee
and
\bee
F = f_{PS} \left[ 1 - \xi (\ln \frac{\mu^2}{m_{PS}^2} +l_4) + \dots \right].
\label{eq:fxifit}
\ee
The $\xi$ parameterization is often used because the
coefficients of higher order terms (here order( $\xi^2$)) are smaller than in the $x$ parameterization.
It is more awkward to use in a fit since the expressions have to be inverted (to express
measured quantities such as $F_{PS} \pm \Delta F_{PS} $ in terms of fit parameters).
 At NLO, the relevant expressions are
\bee
f_{PS} = \frac{F}{2}\left[1 + \left(1+4\frac{m_{PS}^2}{8\pi^2 F^2}(\ln \frac{\mu^2}{m_{PS}^2}  +  l_4) \right)^{1/2} \right]
 \equiv {\cal F}(F, l_4,m_{PS}^2) .
\label{eq:fxi1}
\ee
and
\bee
m_q = \frac{m_{PS}^2}{2B} \left[ 1 + \frac{1}{2}\left( \frac{m_{PS}^2}{8\pi^2 {\cal F}(F, l_4,m_{PS}^2)}\right)
(\log\frac{\mu^2}{m_{PS}^2} +  l_3 ) \right] .
\label{eq:bxifit1}
\ee

 The $U(N_f)$ chiral Lagrangian includes the eta prime in its degrees of freedom.
The appropriate effective field theory was developed in 
Refs.~\cite{DiVecchia:1980yfw,Rosenzweig:1979ay,Witten:1980sp,Kawarabayashi:1980dp,Herrera-Siklody:1996tqr,Kaiser:2000gs}.
The power counting is
\bee
O(\delta) \sim O(p^2) \sim O(M^2) \sim O(m_q) \sim(1/N_c)
\label{eq:deltau}
\ee
(Notice that $1/F^2$ is $O(\delta)$ and  the  $SU(N_f)$ chiral 
expansion parameter $x$ of Eq.~\ref{eq:xxxx} is $O(\delta^2)$.

For $f_{PS}$ and $m_{PS}^2$ the many LEC's of the complete theory
reduce to two, each with its own expansion in $N_c$,
as given by Eq.~\ref{eq:linc} as in the $SU(N_f)$ case.
The factors of $N_c$ in the various $l_i$'s of the $SU(N_f)$ expressions give $O(1/\delta)$ scaling factors
in the $U(N_f)$ perturbative expansion. Thus, all that survives from the $x^2$ 
terms $T_M$ and $T_F$ in Eq.~\ref{eq:tftm} are constants proportional to $N_c^2$.

The eta prime enters as an additional Nambu-Goldstone boson of the non-anomalous 
(in the large $N_c$ limit) singlet $U(1)_A$ symmetry.
It appears in the perturbative expansion as an extra tadpole contribution.
The perturbative calculations for $F_{PS}$ and $M_{PS}^2$ can be found in Ref.~\cite{Guo:2015xva},
are succinctly presented for the degenerate-flavor case in Ref.~\cite{Hernandez:2019qed},
and can be re-assembled (if desired) by slightly editing the calculation of Ref.~\cite{DeGrand:2016pgq}.

The formulas for the chiral expansion need the mass of the eta prime and its decay constant as a function of $N_c$.
The calculation of the eta prime mass for $N_c=3$ is already a 
difficult problem (the statistics of a recent study \cite{Bali:2021qem}
dwarf those of our little project) and so we  
have recourse to theory. Witten and Veneziano \cite{Witten:1979vv,Veneziano:1979ec}
 related the eta prime mass to the quenched
topological susceptibility $\chi_T$ and the pseudoscalar decay constant $F$ (recall the 132 MeV convention),
\bee
M_{\eta'}^2 = M^2 + M_0^2
\label{eq:etaprime}
\ee
where the zero-quark mass eta prime mass is
\bee
M_0^2 =  \frac{4 N_f \chi_T}{F^2}.
\ee
Note that $M_0^2$ is also $O(\delta)$.
There is indirect lattice evidence in favor of this relation from the study of
 the $N_f=2$ topological susceptibility across $N_c$ from Ref.~\cite{DeGrand:2020utq}.
A phenomenological formula which relates the quenched topological
susceptibility to the topological susceptibility at small quark mass due to
Refs.~\cite{DiVecchia:1980yfw},
\cite{Leutwyler:1992yt}
and \cite{Crewther:1977ce} agreed qualitatively with lattice results.

We thus proceed, taking Eq.~\ref{eq:etaprime} for the eta prime mass,
assuming that the eta prime and the pions share a common decay constant, and defining
\beea
x_{\eta'} &=& \frac{M_{\eta'}^2}{8\pi^2 F^2} \nonumber \\
x_0  &=& \frac{M_0^2}{8\pi^2 F^2} \nonumber \\
\eea
to write
\beea
m_{PS}^2 &=&  M^2_{NLO} + M^2_{NNLO} \nonumber \\
f_{PS} &=& F_{NLO} + F_{NNLO} \nonumber \\
\label{eq:uall}
\eea
where
\beea
M^2_{NLO} &=& 2Bm_q\left[ 1- \frac{x}{2}N_cl_3^{(0)} \right] \nonumber \\
F_{NLO} &=& F\left[1 + x N_c l_4^{(0)} \right] \nonumber \\
\label{eq:unlo}
\eea
and
\beea
M^2_{NNLO} &=& 2Bm_q\left[  -\frac{x}{2}( \log\frac{\mu^2}{M^2} + l_3^{(1))}
 +\frac{x_{\eta'}}{2}\log\frac{\mu^2}{M_\eta^2}  +x^2 N_c^2 T_M \right] \nonumber \\
F_{NNLO} &=& F\left[ x(\log\frac{\mu^2}{M^2} +  l_4^{(1)}) + x^2 N_c^2 T_F \right] . \nonumber \\
\label{eq:unnlo}
\eea

We use Eqs.~\ref{eq:uall}-\ref{eq:unnlo} for fits to the data at an individual $N_c$ value.
Notice that, in that case, the NLO fits give $l_i^{(0)}$, but the NNLO fits can only give the full
$l_i= N_c l_i^{(0)} + l_i^{(1)}$. We can present results for the LEC's as we do for the $SU(N_f)$ case, 
a plot of, for example, $F$ versus $1/N_c$.

We can also imagine doing combined fits to the data. In that case we would make an ansatz, that
\beea
F &=& \sqrt{\frac{N_c}{3}}(f_0 + \frac{f_1}{N_c} + \frac{f_2}{N_c^2}) \nonumber \\
B &=& b_0 + \frac{b_1}{N_c} + \frac{b_2}{N_c^2}. \nonumber \\
\eea
We would insert these expressions in Eqs.~\ref{eq:unlo}-\ref{eq:unnlo}, combine them systematically
along with the other LEC's, and fit a range of $N_c$ values.
In order to do this, of course, the $U(N_f)$ chiral expansion has to be well behaved for all the
individual $N_c$ values included in the fit. We return to this point when we present fit results in
Sec.~\ref{sec:results}, but for now we remark that our $N_c=3$ data is incompatible with
the $U(2)$ expansion leaving only two $N_c$'s to work with; too few values of $N_c$ to think about
NNLO fits. So we did not pursue this line.

We note that the LEC's of $SU(N_f)$ chiral perturbation theory and $U(N_f)$ chiral 
perturbation theory are not identical \cite{Kaiser:2000gs,Herrera-Siklody:1998dxd}.
In particular, there are shifts
\beea
B(SU(N_f)) &=& B(U(N_f))\left[ 1 + \frac{1}{N_f} x_0 \log\frac{\mu^2}{M_0^2} \right] \nonumber \\
L_3^{(1)}((SU(N_f)) &=& L_3^{(1)}((U(N_f)) + \log\frac{\mu^2}{M_0^2} -1 \nonumber \\
\label{eq:unsus}
\eea
The shift in the $B's$ is formally order $1/N_c^2$ but the prefactor happens to be large. Note also that,
in contrast to the $SU(N_f)$ case,  in the $U(N_f)$ expansion a shift in the 
regularization point $\mu$ affects the value of $B$. FLAG \cite{FlavourLatticeAveragingGroup:2019iem}
quotes values for the LEC's using $\mu^2$
set to the physical squared pion mass. This choice causes a large shift between the NLO and NNLO
determinations of $B$ at low $N_c$. To avoid this large logarithm, we choose a regularization point closer
to the eta prime mass; we take $\mu^2$ to be an $N_c-$ independent quantity which we choose to
be $\mu^2=8\pi^2 f_{\pi}^2$ where $f_{\pi}=132$ MeV in all our comparison of lattice data
to the $U(N_f)$ expansion.
When we compare the $U(2)$ LEC's to the $SU(2)$ ones, there
 will also be a shift in the $SU(2)$ $l_i's$ due to the different choices for $\mu$,
see Eq.~\ref{eq:switch}.

\section{Details of the numerical simulations \label{sec:latticetech}}

The project involves simulations at three values of $N_c$. Each $N_c$ value requires
its own set of simulations
at four bare gauge couplings and 4-6 bare quark masses per gauge coupling.
At each bare parameter value six
quantities are measured: the Axial Ward Identity fermion mass, the pseudoscalar mass,
 the pseudoscalar decay constant,
two lattice to continuum renormalization factors (for the fermion mass and decay constant),
as well as a parameter to set the lattice spacing. All these quantities must be checked for correlations.
The lattice regulated quantities are converted into dimensionless continuum regulated quantities which are
then fit to the formulas described in the last section, plus lattice artifacts.
Doing this is a long and somewhat recursive task.
We believe that the methodology we use to generate, collect, and analyze our lattice data sets
is completely standard. However, all lattice calculations involve many choices. This section
 describes what we did in perhaps too much detail. 

\subsection{Simulation details: lattice action, updating algorithm, data sets, observables}
Our simulations are done with the usual Wilson plaquette action, with the bare gauge 
coupling $g_0$ labelled by  $\beta = 2N_c / g_0^2$.
Two flavors of degenerate mass fermions (discretized as Wilson-clover fermions)
are included.
Configurations are generated using
 the Hybrid Monte Carlo (HMC)  algorithm \cite{Duane:1986iw,Duane:1985hz,Gottlieb:1987mq}
with a multi-level Omelyan integrator \cite{Takaishi:2005tz} and
multiple integration time steps \cite{Urbach:2005ji}
with one level of mass preconditioning for the fermions \cite{Hasenbusch:2001ne}.

The fermion action uses gauge connections defined as normalized hypercubic (nHYP)
 smeared links~\cite{Hasenfratz:2001hp,Hasenfratz:2007rf,DeGrand:2012qa}.
Simulations use the
arbitrary $N_c$ implementation of Ref.~\cite{DeGrand:2016pur}.
The bare quark mass $m_0^q$ appears in the lattice action along with the lattice spacing $a$
 via the hopping
parameter $\kappa=(2m_0^q a+8)^{-1}$. The clover coefficient is fixed to
 its tree level value, $c_{\text{SW}}=1$.
 The gauge fields obey periodic
boundary conditions; the fermions are periodic in space and antiperiodic in time.

Lattice volumes are a mix of $16^3\times 32$ and $24^3\times 32$  sites. The lattice sizes were chosen so
as to minimize finite volume effects (which we describe in Sec.~\ref{sec:finitev} below).
Briefly, we began simulation runs at nearly all our bare parameter values taking $16^3$ spatial
volumes and collected enough data to determine the mass  and decay constant of a pseudoscalar meson, 
and then to estimate the size of the finite volume correction. When this appeared to be
larger than a few per cent, we replaced the $16^3$ data sets by $24^3$ ones. A glance
at  Tables~\ref{tab:data3}-\ref{tab:data5} shows
that there are more $24^3$ $N_c=3$ data sets than there are for $N_c=4$ or 5. This is because of the scaling
of the pseudoscalar decay constant with $N_c$, as  described in the Introduction.

Lattices used for analysis are spaced a minimum of 10 HMC time units apart,
so individual bare parameter sets
contain 30 to 200  stored lattices.
All data sets at individual bare parameters $(\beta,\kappa)$  are based on a single stream.
We check for thermalization at the start of each data stream (which typically begins from an equilibrated
configuration at a nearby bare parameter set) and typically discard the first 100 trajectories (more than this
for simulations at strong coupling).
The data sets are extensions of ones presented in Refs.~\cite{DeGrand:2016pur,DeGrand:2017gbi}.

Data for the pseudoscalar mass, the pseudoscalar decay constant, and the fermion mass
 come from hadron correlators using
propagators constructed in Coulomb gauge, with Gaussian sources and
 $\vec p=0$ point sinks. We tune the width of the source to minimize the dependence
of the effective mass
(defined, as usual, to be $m_{eff}(t) = \log C(t)/C(t+1)$ in the case of open boundary conditions for the
 correlator $C(t))$ on the distance $t$ between source and sink.
 All results are based on a standard full correlated
analysis involving fits to a wide range of $t$'s.
Best fits are chosen with the ``model averaging'' ansatz of Jay and Neil \cite{Jay:2020jkz}.
In one sentence, this gives each particular fit in a suite (with a chi-squared value $\chi^2$ and
$N_{DOF}$ degrees of freedom) a weight in the average which is proportional to
$\exp(-(\chi^2/2 - N_{DOF}))$.

The pseudoscalar decay constant is
defined in terms of the matrix element of the timelike component of the axial current
$A_0(x,t)=  \bar u(x,t) \gamma_0 \gamma_5 d(x,t)$ between the vacuum and a pseudoscalar state as in Eq.~\ref{eq:fpi}.
 It is determined via a correlated fit to two propagators
with a common source,
\beea
C_1(t)=\sum_\bx  \svev{ A_0(\bx,t)\co(0,0)} \nonumber \\
C_2(t)=\sum_\bx  \svev{ \co(\bx,t)\co(0,0)}. \nonumber \\
\eea

The fermion mass  $am_q$  is taken to be the Axial Ward Identity (AWI) fermion mass, 
defined through the relation of the axial current matrix element to the pseudoscalar
$P(x,t)= \bar u(x,t) \gamma_5 d(x,t)$,
\bee
\partial_t \sum_\bx \svev{A_0(\bx,t)\co} = 2am_q \sum_\bx \svev{ P^a(\bx,t)\co},
\label{eq:AWI}
\ee
again taken from a simultaneous fit to two two-point functions.
$\co$ can be any convenient source, of course, and for us it is a Gaussian.

Conversion factors from lattice regulated quantities will be needed for $f_{PS}$ and $m_q$ and 
their calculation will be described in Sec.~\ref{sec:zfactors} below.

The pseudoscalar mass $am_{PS}$  is determined from fits involving correlators of pseudoscalar
currents.

We give is a brief summary of the data sets. We attempted to collect data at similar
 values of $\xi=m_{PS}^2/(8\pi^2 f_{PS}^2)$ for our different values of $N_c$,
since that is the expansion parameter for chiral perturbation theory.

For $N_c=3$ we have four  beta values and 24 bare parameter sets in total. The $\xi$ range is from
0.06 to 0.19. $t_0 m_{PS}^2$ ranges from 0.05 to 0.35, for pion masses in
 the range 300 - 776 MeV (taking a nominal value for the flow parameter
$\sqrt{t_0}=0.15$ fm from Ref.~\cite{Sommer:2014mea}). $t_0/a^2$ ranges from
1.0 to 2.4 for lattice spacings
between 0.10 - 0.15 fm.

The $N_c=4$ data sets come from four beta values and there are 26 bare parameter sets. $\xi$ ranges from 0.04 to 0.2
and $t_0m_{PS}^2$ is in the range 0.06-0.67.
$t_0/a^2$ is in the range 1.1-3.34 for $a=0.08-0.14$ fm.

And there are four beta values and 21 bare parameter sets in the $N_c=5$ sector. $\xi$ ranges from 0.05 to 0.16,
$t_0 m_{PS}^2$ is in the range 0.10-0.63, while $t_0$ lies between 1.5 and 3.4, or $a=0.08-0.12$ fm.

A useful marker point is $\xi=0.1$, for which $t_0 m_{PS}^2$ is about 0.10, 0.22, and 0.35 for $N_c=3$, 4, or 5.
This quantity will re-appear in Sec.~\ref{sec:results}.

Another  marker is a comparison with the  large $N_c$ study of Ref.~\cite{Hernandez:2019qed}.
Their data sets were collected at fixed fermion masses across $N_c$ and spanned
$\xi \in (0.10-0.16)$ for $N_c=3$,
$\xi \in (0.05-0.09)$ for $N_c=4$,
$\xi \in (0.04-0.08)$ for $N_c=5$, and
$\xi \in (0.03-0.06)$ for $N_c=6$. The lower part of our $\xi$ ranges overlap with theirs.

The data are presented in Tables~\ref{tab:data3}-\ref{tab:data5}.

\subsection{Setting a scale\label{sec:t0}}

The conversion of simulations quantities whose scale is set by the lattice spacing
into dimensionless parameters which can be used in continuum extrapolations
is done using the Wilson flow parameter
 $t_0$ \cite{other,Luscher:2010iy}.
The determination of $t_0$ from each data set (with its own $\beta$ and $\kappa$)
is done in the standard way, from the observable
$E(t_0)$  extracted from the field strength tensor,
\bee
t_0^2 \svev{E(t_0)} = C(N_c)  .
\label{eq:flow}
\ee
 $C(N_c)$ is chosen to match what most other large $N_c$
simulations take,
\bee
 C(N_c)= C \left( \frac{3}{8} \frac{N_c^2-1}{N_c}\right),
\label{eq:ce}
\ee
with $C=0.3$  the usual value used in  $SU(3)$.
Our procedure for computing
 $t_0$ from the data is identical to what was done in Ref.~\cite{DeGrand:2020utq}
and details may be found there.

In contrast to our earlier work, where conversion from dimensionful to dimensionless parameter
($(am_{PS})^2$ to $t_0 m_{PS}^2$, for example) was done at individual bare parameter values
with a $t_0(\beta,\kappa)$, we need to use a mass-independent definition of $t_0$.
This is because the chiral expansion for $t_0$ has its own set of NLO analytic contributions
which combine with the ones for our desired observables ($f_{PS}$ and $m_{PS}^2$) \cite{Bar:2013ora}
and would contaminate the chiral LEC's. We also need a scale choice which is superficially identical across $N_c$.

We investigated several possibilities
\begin{itemize}
\item $t_0$ at a fixed value $(N_c/3)\xi$ -- This is a bit awkward to produce, since it needs $Z_A$, 
the axial current matching factor.
\item $t_0$ at fixed $m_{PS}^2/m_V^2$ (where $m_V$ is the mass of the vector meson) -- 
This was noisier than the first choice.
\item $t_0$ at fixed  $t_0m_{PS}^2$ -- This was also noisier than the first choice (and with greater correlations).
\end{itemize}
We arbitrarily selected the first choice, $(N_c/3)\xi \equiv \xi_0$.

Then we followed the following procedure to produce a value of $t_0$ at each $\beta$ value.
Each data set gives it own $\xi_0 \pm \Delta \xi_0$ and $t_0 \pm \Delta t_0$.
For each $\beta$ value, we
performed a polynomial fit $t_0(i) = \sum_n C_n (\xi(i)-\xi_0)^n$ to produce a value of $t_0(\xi_0,\beta)$.
The fit included the
 errors $\Delta t_0$ and $\Delta \xi$. (See Sec.~\ref{sec:fitting} for a discussion about fit methodology.)
We checked subsets of the data to see if correlations in $t_0$ and $\xi$ affected the fit. They did not
(differences in correlated and uncorrelated fits for 
 $t_0(\xi_0)$ were well under a standard deviation) so we did not include them in our analysis.
We typically fit to a cubic, bracketing $\xi_0$ so that the fit was essentially an interpolation.

Our choice of $\xi_0$ is purely heuristic. We have to make
 sure $\xi_0$ lies inside the range of measured $(N_c/3)\xi$ values
recorded for each bare parameter set, and then we simply pick a $\xi_0$  for which $\Delta t_0$ is minimized.
Like many other decisions made in the data analysis, we looked at many possible choices
 but we saw little effect on results. In the end, we took $\xi_0=0.12$.

\begin{figure}
\begin{center}
\includegraphics[width=0.8\textwidth,clip]{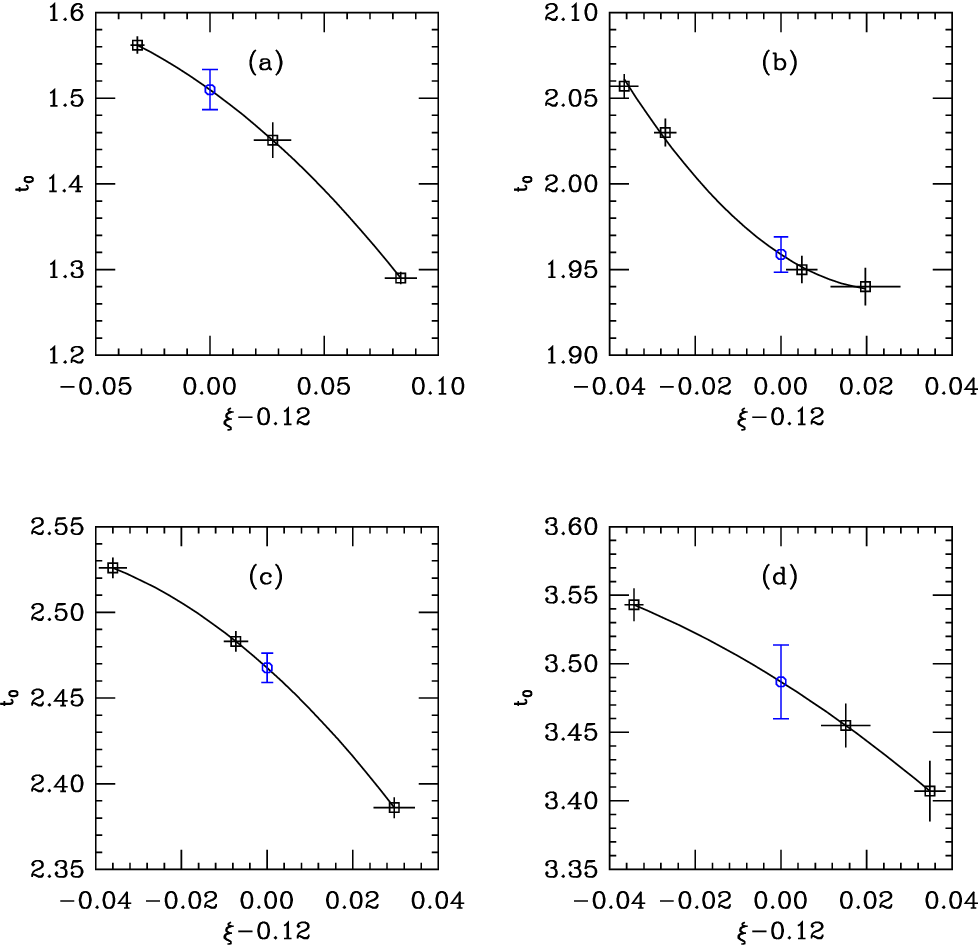}
\end{center}
\caption{Interpolating to find a fiducial $t_0$, for $N_c=5$ data sets. 
The data at individual kappa values are shown in squares;
the interpolated value is an octagon (in blue). Panels are
(a) $\beta=16.2$,
(b) $\beta=16.3$,
(c) $\beta=16.4$
(d) $\beta-16.6$.
\label{fig:t0su5}}
\end{figure}

An example of the extraction of $t_0$ from data, for $N_c=5$, is shown in Fig.~\ref{fig:t0su5},
and the data are collected in Table~\ref{tab:t0}.

\subsection{Correcting for finite volume\label{sec:finitev}}
Our smallest quark mass data could potentially be affected by our simulation volumes.
For chiral observables, the cause is well known: tadpole graphs,
where a pseudoscalar particle  is emitted from some location
and returns to the same point,
 are replaced by a set of contributions connecting the location to its image points.
(We found the discussion in Ref.~\cite{Sharpe:1992ft} to be quite helpful.)
If we  write the pseudoscalar correlator for a particle of mass $m$
 in a box of length $L_\mu$ in direction $\mu$ as
\bee
\Delta(m,x) \rightarrow \sum_{n_\mu} \Delta(m,x+n_\mu L_\mu),
\label{eq:fs}
\ee
the infinite volume propagator, call it $\bar \Delta(m,x)$, is the $n=0$ term in the sum.
The finite volume tadpole is
\bee
\Delta(m,0)= \bar\Delta(m,0) + \bar I_1(m,L)
\ee
where $\bar I_1(m,L)$ is the sum over images.
If a typical infinite volume observable has a chiral expansion
\bee
O(L=\infty)=O_0[1+ C_0 \frac{1}{f_{PS}^2} \bar \Delta(m,0) ]
\label{eq:typical}
\ee
then the finite volume correction is
\bee
O(L)-O(L=\infty)=O_0[ C_0 \frac{1}{f_{PS}^2} \bar I_1(m,L) ].
\ee
We replace the simulation observable in our data set $O(L)$ by
\bee
O(L=\infty)= \frac{O(L)}{1+C_0 \frac{1}{f_{PS}^2} \bar I_1(m,L)} .
\label{eq:fvcorr}
\ee
$\bar I_1(m,L)$ can be found from
\beea
\Delta(x) &=& \int\frac{d^4 p}{(2\pi)^4}\frac{\exp(ipx)}{p^2+m^2} \nonumber \\
&=& \frac{m}{16\pi^2 x}\int_0^\infty dk \exp(-mx(\frac{k}{4}+\frac{1}{k})) \nonumber \\
\eea
by summing over all image points until the expression saturates.

Eq.~\ref{eq:fvcorr} also needs $f_{PS}$,  the decay constant in the chiral limit.
This is, of course a parameter in the fit. However, if we just make an estimate for it
and evaluate Eq.~\ref{eq:fvcorr}, we find that for all our data sets, the correction is
at most on the order of a few per cent for a few of the lightest-mass points, so we only need to input an approximate value 
of $f_{PS}$ given all the other uncertainties in the calculation.
We use $\sqrt{t_0}f_{PS}=0.078$, 0.107, 0.128 for $N_c=3$, 4, 5.

\subsection{Lattice to continuum regularization conversion\label{sec:zfactors} }

We will use the ``regularization independent'' or RI scheme \cite{Martinelli:1994ty}
for  computing matching
factors (labelled as $Z_i$ for current $i$).
The method is standard and so we will only briefly describe our particular implementation of it.

 One computes quark and gluon Green's
functions in Landau gauge, regulated by giving all external lines a
large Euclidean squared momentum $p^2 = \mu^2$, and uses combinations of them
to determine the $Z$'s.
Specifically, we define
\bee
\mu^2 = \sum_i 4 \sin^2 \frac{p_i}{2} + 4 \sin^2 \frac{p_t}{2}
\ee
where $p_i=2n_i\pi/L$ and $p_t=(2n_i+1)\pi/T$ since the temporal boundary conditions are
anti-periodic. (Recall that the lattices are $L^3\times T$ sites.)

The matching factor is defined by computing the ratio of the matrix
elements of the operator between off-shell single particle momentum
eigenstates in the full and free theories
\bee
Z^{RI}_\Gamma(\mu) \langle p| O_\Gamma|p\rangle_{p^2=\mu^2} =
  \langle p|O_\Gamma|p\rangle_0  .
\ee
or equivalently
\bee
Z^{RI}_\Gamma(\mu) \frac{1}{4N_c} \Tr[ \langle p| O_\Gamma|p\rangle_{p^2=\mu^2}
  \langle p|O_\Gamma|p\rangle^{-1}_0] =1 ,
\ee
where the factor of $1/4N_c$ counts Dirac spins and colors.
We use the $RI^\prime$ scheme to
define $Z_Q$ from a projection against the free propagator $S_0(p)$:
\bee
Z^{RI^\prime}_Q = \frac{1}{4N_c} \Tr S(p) S_0^{-1}(p).
\ee

We are using clover fermions, for which there is an overall $2\kappa$ factor in the field definition as opposed
to a continuum-like normalization. We define our $Z_\Gamma$ factors for
observables built from clover fermions   by
\bee
 \svev{h|\Gamma|h'}_{cont} = \svev{h|\Gamma|h'}_{latt}\times 2\kappa Z_\Gamma
\ee
where $Z_\Gamma \sim 1+C g^2...$ in a perturbative expansion. The $2\kappa$ factor gives clover
Z's equal to unity for free field theory.

One needs the scale $\mu$  to be large enough not to be affected by confinement effects but not so large
that is is affected by lattice artifacts. Given our lattice spacings and volumes, this means in practice
a value of $\mu a$ around unity.

We need a renormalization factor for the axial current, $Z_A$, to define $F$ and one for the fermion mass $Z_m$
to carry the quark mass to a continuum regularization.
  The computations are straightforward. We generate lattice data for $Z_A(a\mu)$ at many
values of $(a\mu)^2 = \sum_j a^2 p_j^2$ for lattice momenta $ap_j$.
The data is averaged under single elimination jackknife and sorted in $a\mu$.
Fits to a linear function are performed over ``windows'' of data around a central value and checked
for systematic dependence on the chosen central value.
We typically use about 30 lattices per bare parameter value to do this. The resulting uncertainties
are completely dominated by systematics, to be described below,
so it is not worthwhile to work with larger data sets.

\subsubsection{$Z_A$}

\begin{figure}
\begin{center}
\includegraphics[width=0.8\textwidth,clip]{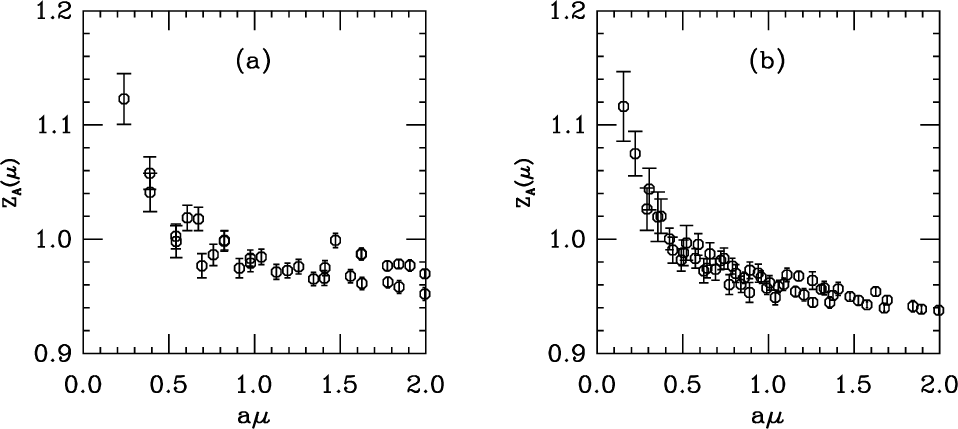}
\end{center}
\caption{ Examples of data sets and results for $Z_A$. Panel (a)
is from the $\beta=5.35$, $\kappa=0.12775$, $L=16$ set
and (b) from $\beta=5.35$, $\kappa=0.1282$, $L=24$.
They show $Z_A$ versus $a\mu$ around $ a\mu=1$.
\label{fig:za1624}}
\end{figure}

 Fig.~\ref{fig:za1624} illustrates our results for $Z_A$. It shows large scale views of $Z_A$ for
a $16^3$ and a $24^3$ spatial volume lattice. The $16^3$ volume shows clear discretization artifacts above
$a\mu \sim 1.4$. This represents an upper limit on the range of $a\mu$ values which can be analyzed.
The finer grained $24^3$ data do not have this issue.

Fits about some central value of $a\mu$ typically show a value close to unity with a statistical
uncertainty of a few parts per mille. That $Z_A$ is close to unity for the fermion action used here (nHYP) fermions
has been well documented in the literature (cf. Ref.~\cite{DeGrand:2005af}).

However, there is another issue. $Z_A$ should be a constant across $a\mu$ because
the operator has no anomalous dimension.  Our data sets show a small but noticeable
 drift of $Z_A(a\mu)$ versus $a\mu$.  We account for this drift by comparing linear fits
to  $Z_A(a\mu)$ versus $a\mu$ over  windows of $a\mu$. We pick (somewhat arbitrarily)
a central value $a\mu =1$ and a range $a\mu=0.8- 1.2$.
(For comparison, $\mu=2$ GeV corresponds to $a\mu \sim 1.4$ at our coarsest lattice spacings
down to 0.8 at our finest spacings.) Our systematic uncertainty is taken to be
$|Z_A(a\mu=1.2) - Z_A(a\mu=0.8)|/2$. This gives a systematic uncertainty in the range of 0.005-0.008
as compared to statistical uncertainties from $Z_A(a\mu=1)$ of 0.002-0.003.
We use the first quantity as the uncertainty for $Z_A$.

\subsubsection{$Z_S$ and $Z_m$}

$Z_m$ is a bit more complicated to compute.
 We take $Z_m$ from the renormalization factor for the scalar current $Z_S^{RI}$
as $Z_m= 1/Z_S$, defined at some value of $\mu a$. We must then convert it to an
 $\overline{MS}$ value at a nominal scale $\mu=2$ GeV. (This is done to make contact with
pre-existing $SU(3)$ results.)
This involves two steps after  $Z_S^{RI}(a\mu)$ is determined; the analysis to determine this quantity is
identical to what is done for $Z_A$.
 First, there is the conversion to $\overline{MS}$ at the scale at which $Z_S^{RI}$ is determined.
An explicit three loop formula
 (for general $N_c$ and $N_f$ fundamental representation fermions) for this conversion is
 given by Chetyrkin and Retey \cite{Chetyrkin:1999pq}.
 (See also Ref.~\cite{Franco:1998bm}.) To use it we need the $\overline{MS}$ coupling
at scale $\mu = (\mu a)/a$. The lattice spacing is given by the flow parameter in lattice units
($t_0^L = t_0/a^2$ where $t_0=0.15$ fm from the compilation in Ref.~\cite{Sommer:2014mea}).
The $\overline{MS}$ coupling comes from the plaquette, where in the ``alpha-V'' scheme \cite{Brodsky:1982gc,Lepage:1992xa}
\bee
\ln\frac{1}{N_c}\Tr U_p= -4\pi C_F \alpha_V(q^*_V),
\ee
where $q^*=3.41/a$ for the Wilson plaquette gauge action and $C_F=(N_c^2-1)/(2N_c)$ is the usual
invariant. Then the conversion to $\overline{MS}$ is given by
\bee
\alpha_{\overline{MS}}(e^{-5/6}q^*)=\alpha_V(q^*)(1- \frac{2}{3}N_c \frac{\alpha_V}{\pi})
\label{eq:vtomsbar}
\ee
and run to $\alpha_{\overline{MS}}$(2 GeV) with the usual two-loop formula
\bee
\alpha_s(q) = \frac{\alpha_s(q_0)}{v}(1+\beta_1 \frac{\alpha_s(q_0)}{v}\log v )
\ee
with
\bee
v=1 - \beta_0 \frac{\alpha_s(q_0)}{2\pi}\log \frac{q_0}{q}  .
\ee
In all these equations
 $\beta_0$ and $\beta_1$ are the two lowest coefficients of the beta function.
Recalling that $\beta_0= 11/3N_c - 2/3 N_f$ and $\beta_1=34/3N_c^2-N_f/2((20/3)N_c+4 C_F)$,
we note the appearance of the `t Hooft coupling $N_c \alpha_s$ in all these formulas.
(The $RI$ to $\overline{MS}$ formula, which we do not quote, involves $\alpha_s C_F$ as well.)
We did not find Eq.~\ref{eq:vtomsbar} for $N_c\ne 3$ in the literature but it can be
reconstructed from expressions in Ref.~\cite{Billoire:1979ih}.

Finally, the $\overline{MS}$ $Z_m$ is run to the final scale $\mu=2$ GeV; this is done with the usual
two loop formula
\bee
m_q(q)= m_q(\mu) \left( \frac{\alpha_s(q)}{\alpha_s(\mu)}\right)^{\frac{\gamma_0}{2\beta_0}}
[1 + \frac{\alpha_a(q)-\alpha_s(\mu)}{4\pi}(\frac{\gamma_1 \beta_0 - \beta_1 \gamma_0}{2\beta_0^2}) ] .
\ee
This is always a small effect since $\mu a \sim 1$ puts $\mu$ in close vicinity to 2 GeV already.
 The $\gamma_i$'s are the one and two loop terms for the mass anomalous dimension.

The need to match and run makes the determination of $Z_m$ much more fraught that the case of $Z_A$.
The issue can be seen in the $RI$ to $\overline{MS}$ conversion factor. (Here we describe the situation for $N_c=3$.)
Working at $a \mu \sim 1$, the scale $\mu$ ranges from about 1.4 to 2 GeV and $\alpha(\mu)$ is in the range
0.18-0.2. The conversion factor can be written as
\bee
R = 1 + \sum_i C_i \alpha_s^i.
\ee
The fraction of $R$ from the highest order term, $C_3 \alpha_s^3/R$, is about two per
 cent (slightly bigger at smaller lattice spacing, slightly smaller at bigger lattice spacing, of course).
 This issue has already been noticed and described by  Chetyrkin and Retey \cite{Chetyrkin:1999pq}. The
suggested cure,
to match at bigger $a\mu$, is not possible for our data sets due to the discretization effects we have already described.
We must include  a systematic effect in the error budget. We elect to do this by taking
$C_3 \alpha_s^3/R$ as a fractional systematic uncertainty on $m_q^{\overline{MS}}$.
(Compare the discussion in Ref.~\cite{Budapest-Marseille-Wuppertal:2013vij}, from a study at much smaller lattice spacing.)
We again bracket the match between $a\mu=0.8$ and 1.2.

\begin{figure}
\begin{center}
\includegraphics[width=0.8\textwidth,clip]{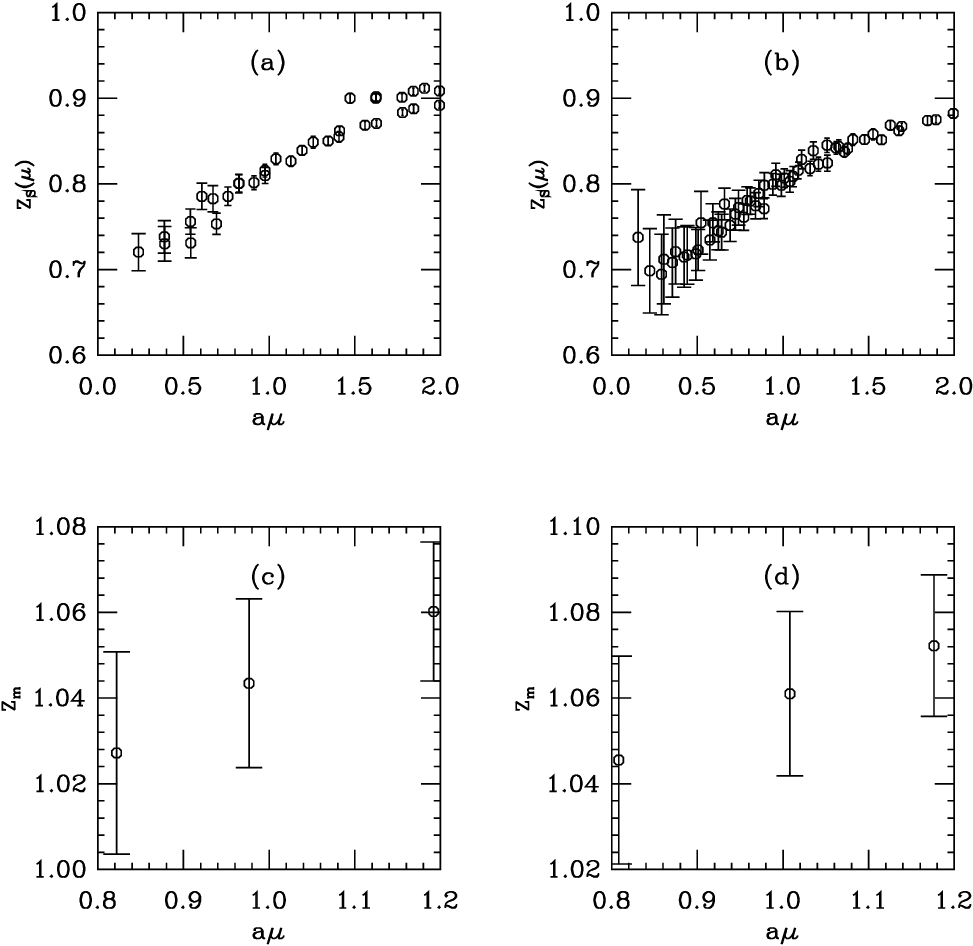}
\end{center}
\caption{ Examples of data sets and results for $Z_S$ and $Z_m$. Panels (a) and (c)
are from $\beta=5.35$, $\kappa=0.12775$, $L=16$ set
and (b) and (d) from $\beta=5.35$, $\kappa=0.1282$, $L=24$.
Panels (a) and (b) show $Z_S$ from lattice data versus $a_\mu$ around $\mu=1$ while panels (c) and (d) show
$Z_m^{\overline{MS}}$ from linear fits to $Z_S(a\mu)$ centered at the shown $a\mu$ values, then
converted into $Z_m^{\overline{MS}}(\mu=2$ GeV$)$.  The error bar includes the factor $C_3 \alpha_s^3/R$, as a systematic.
\label{fig:zs1624}}
\end{figure}

We conclude with some figures showing the analysis for $Z_m$, in Fig.~\ref{fig:zs1624}. The errors on
$Z_m(\mu)$ are $C_3 \alpha_s^3$ term.

One positive remark about $Z_m^{\overline{MS}}$ worth mentioning: the points shown at different $a\mu$ involve
slightly different amounts of correction (since the conversion from $RI'$ to $\overline{MS}$ is done at
 different $a\mu$ values with
different running distances to $\mu=2$ GeV). Nevertheless, the results are reasonably independent
of $a\mu$ even discounting the overall systematic uncertainty.

\section{Fitting strategies\label{sec:fitting}}

We believe that little in our fitting strategy is novel.
Fits involve minimizing a chi-squared function
\bee
\chi^2 = \sum_{ij} X^T(i) \frac{C(i,j)^{-1}}{\Delta E(j) \Delta E(j)} X(j),
\label{eq:chisq}
\ee
where the sum runs over all quantities determined by simulation and the entries are
$X(i)= E(i) - T(i)$ where $E(i)$ is the measured data value, $\Delta E(i)$ is
 its standard deviation of the mean, $T(i)$ is the model (containing parameters to be fit)
and $C(i,j)$ is the correlation between the different experimental values. If the data points
are assumed to be uncorrelated, $C(i,j)^{-1} = \delta(i,j)$ in our convention.

We include correlations in  Eq.~\ref{eq:chisq} by jackknife. Recall that under a jackknife the average
a quantity $\overline x$ and its uncertainty $\Delta x$ are given in terms of
$N$ individual jackknife averages
$x(n)$
\beea
\overline x &=& \frac{1}{N}\sum_n x(n) \nonumber \\
(\Delta x)^2 &=&  \frac{N-1}{N}\frac{1}{N}\sum_n (x(n)^2 - \overline x^2). \nonumber \\
\eea
The correlation function is computed similarly, with a convention that the diagonal entries are the identity:
\bee
C(i,j)= \frac{1}{\Delta x_i \Delta x_j} \frac{N-1}{N} \sum_n (x_i(n)-\overline x_i)(x_j(n)-\overline x_j).
\ee

Recall all the ingredients we need: $am_q$, $am_{PS}$, $af_L$, $t_0/a^2$, $Z_A$, $Z_m$.
We combine these into dimensionless continuum-regulated quantities, which are then used as input data
to fits which determine the LO and NLO chiral parameters.
The dimensionless quantities we need, compulsively retaining the lattice spacing, are
\beea
\hat m_q &=& Z_m \sqrt{\frac{t_0}{a^2}} am_q \nonumber \\
\hat m_{PS}^2 &=& \frac{t_0}{a^2} (am_{PS})^2  \nonumber \\
\hat f_{PS} &=& Z_a \sqrt{\frac{t_0}{a^2}} af_{PS} . \nonumber \\
\eea
In practice, only $am_q$, $am_{PS}$ and $af_L$
show measurable correlations and we retain only these correlations in the chiral fits.
The uncertainties on $\hat m_q$, $\hat f_{PS}$, and $\hat m_{PS}^2$ come from the uncertainties in their ingredients,
combined in quadrature. We also need $t_0/a^2$ (actually, its inverse)
 as a (squared) lattice spacing to add lattice dependent terms in
 the chiral fits.

In Sec.~\ref{sec:t0} we mentioned the issue that we have to perform
a fit determining a set of parameters $\{a\}$ from  $y(i)=f(x_i,\{a\})$, where both the $y_i$'s and $x_i's$
have uncertaintities. We deal with that issue by promoting the $x_i$'s in the fitting function to additional
terms in the $\chi^2$ function. For example, in Sec.~\ref{sec:t0} the $y_i$'s were values of
 $t_0\pm\Delta t_0$ at a set of
$N$ bare parameter values, and the $x_i$'s were a set $\xi_i\pm \Delta \xi_i$. The set  $\{a\}$ were a set 
of $J$ coefficients on a polynomial fit. We expand the $\chi^2$ function to
\bee
\chi^2 = \sum_{i=1}^N \frac{(y_i - f(X_i,\{a\}))^2}{(\Delta y(i))^2} + \sum_{i=1}^N \frac{(x_i - X_i)^2}{(\Delta x(i))^2}
\ee
(we suppress the obvious correlation term), so that we now have $2N$ terms in $\chi^2$ and $J+N$ 
quantities to be determined. If there were no errors on the $x_i$  
the counting of degrees of freedom would  be $N-J$,  and with the procedure we have outlined it is still
$2N - (J+N) =N-J$.

The $\xi$ fits present one final annoyance, since the fitting function must be written entirely in
 terms of quantities to be fit.
This means that in Eq.~\ref{eq:bxifit1} we must
remove the error-bearing $m_{PS}^2$ from the right hand side of these expressions. We do this by adding
 a  $ (y(i)-T(i))=(m_{PS}(i)^2-M_{PS}(i)^2)$ term to the chi-squared formula.
For $N_d$ bare parameter values there  will be $3N_d$ data points
($F_{PS}$, $m_q$ and $M_{PS}$) to be fit, with $4+3N_d$ fit parameters, for $N_d-4$ degrees of freedom.

Finally, we must discuss how we deal with lattice artifacts. 
Our lattice action has order $a^2$ artifacts, and this
means that the LEC's which will come out of fits will also carry $O(a^2)$ corrections.
In principle, for $x$ fits, that will be the case for all four LO and NLO parameters and the three additional NNLO ones.
To include all these correction terms in the fitting function would be very unwieldy (and, given
our uncertainties, quite unstable). Instead, we follow
the lead of Ref.~\cite{Budapest-Marseille-Wuppertal:2013vij} and proceed empirically, adding terms
 and seeing how they affect
the $\chi^2$ of a fit.
We discover that we are sensitive to lattice artifacts in two places, basically a modification 
of the right hand sides of Eqs. \ref{eq:xfits}, \ref{eq:fxi1} and \ref{eq:bxifit1}.
To be explicit, we fit the following functional forms. For $x$ fits,
\beea
m_{PS}^2 &=& (1+c_B \frac{a^2}{t_0}) 2Bm_q \left[ 1 - \frac{1}{2}x(\log\frac{\mu_\pi^2}{M^2} + l_3) + x^2(\frac{17}{8}T_M^2 + k_M) \right] \nonumber \\
f_{PS} &=& (1+c_F \frac{a^2}{t_0})   F  \left[ 1 + x(\log\frac{\mu_\pi^2}{M^2} + l_4) + x^2(-\frac{5}{4}T_F^2 + k_F) \right]  , \nonumber \\
\label{eq:realxfit}
\eea
$T_F$ and $T_M$ are given by Eq.~\ref{eq:tftm}.  The NLO $\xi$ fitting functions become
\beea
f_{PS} &=&  (1+c_F\frac{a^2}{t_0}) \frac{F}{2}\left[1 + \left(1+4\frac{m_{PS}^2}{8\pi^2 F^2}(\ln \frac{\mu^2}{m_{PS}^2}  +  l_4) \right)^{1/2} \right] \nonumber \\
m_q &=&  (1+c_B\frac{a^2}{t_0})\frac{m_{PS}^2}{2B} \left[ 1 + \frac{1}{2}\left( \frac{m_{PS}^2}{8\pi^2 {\cal F}(F, l_4,m_{PS}^2)}\right)
(\log(\frac{\mu^2}{m_{PS}^2}) +  l_3 ) \right] . \nonumber \\
\label{eq:realxifit}
\eea
Note again that all input variables ($m_q, M_{PS}^2, F_{PS}$) have been rescaled by the appropriate power of $t_0$.
This means that  an NLO fit involves six parameters and NNLO, nine. Of these two nuisance parameters,
$c_B$ is much more important. (It was the only term kept by Ref.~\cite{Budapest-Marseille-Wuppertal:2013vij}.)
We comment on their values in the next section.

\begin{figure}
\begin{center}
\includegraphics[width=0.7\textwidth,clip]{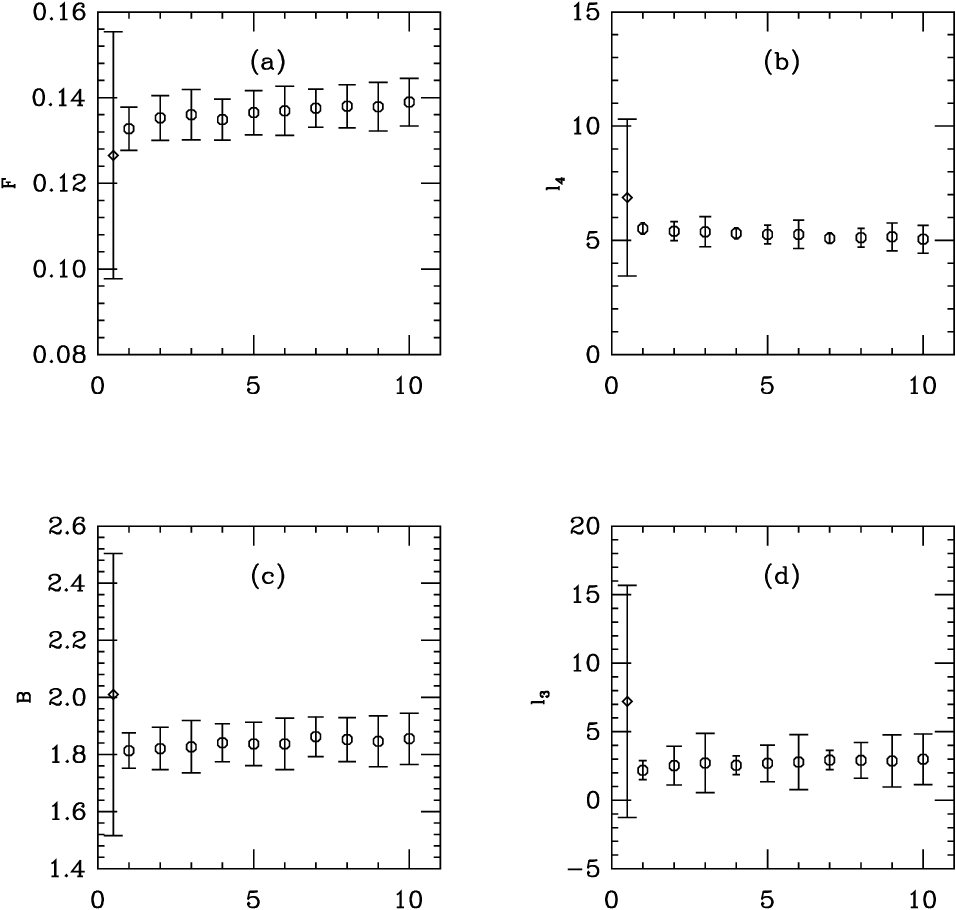}
\end{center}
\caption{Fit values of LO and NLO chiral parameters from an NNLO $x$ fit  to $N_c=5$ over a set of priors
for $l_{12}$, $k_F$, and $k_M$. The diamond shows the fit result without priors.
\label{fig:priorsu5}}
\end{figure}

\begin{figure}
\begin{center}
\includegraphics[width=0.7\textwidth,clip]{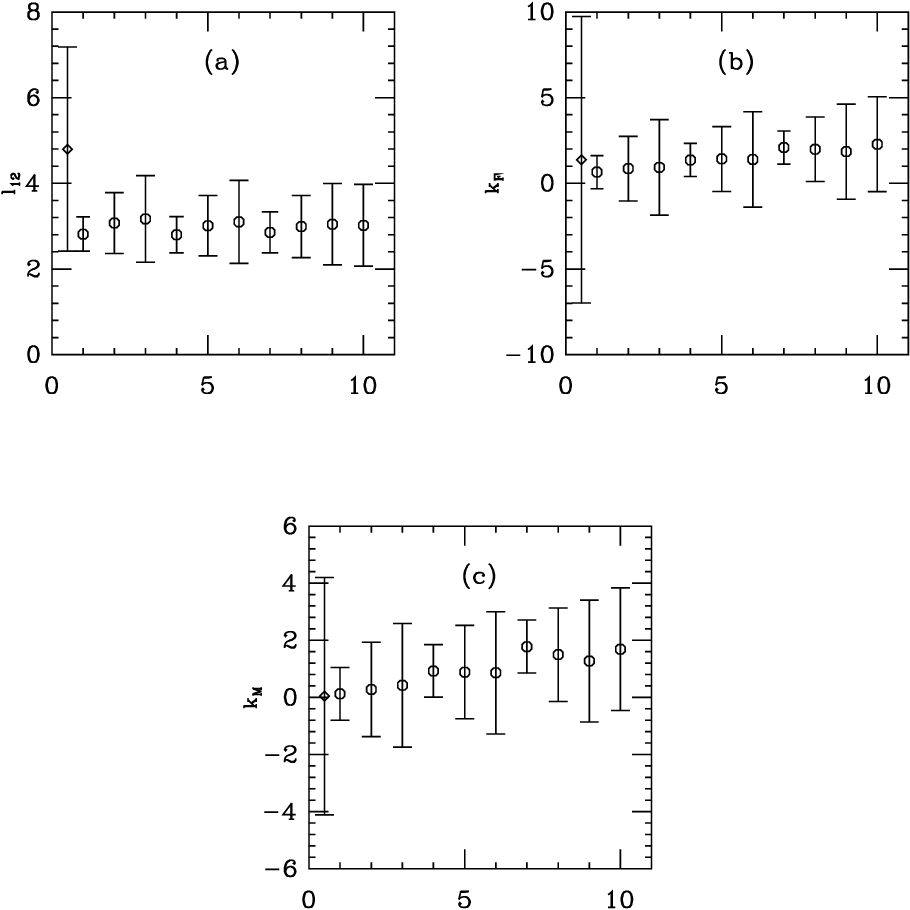}
\end{center}
\caption{Fit values of NNLO parameters $l_{12}$, $k_F$, and $k_M$   from
NNLO $x$  fits to $N_c=5$ over a set of priors for them.
 The diamond shows the fit result without priors.
\label{fig:prior2su5}}
\end{figure}

Priors are included in the standard way, as additional terms
in the $\chi^2$ function, $\chi^2 \rightarrow \chi^2 + \chi_P^2$ where
\bee
\chi_P^2 = \sum_j\frac{(p_j-P_j)^2}{\Delta P(j)^2}.
\ee
We discover that while both $x$ and $\xi$ NLO fits are uniformly stable over a wide range of choices of
fit parameters, the NNLO fits are unstable without inclusion of priors.
 We made a set of fits with a broad range of priors for $l_{12}$, $k_F$, and $k_M$, and we found  that 
values of the four fitted LO and NLO
quantities are  reasonably independent of the choices we made, but the fitted values and (especially) their uncertainties
of $l_{12}$, $k_F$, and $k_M$
are completely determined by the priors. This result occurs for all $N_c$'s. We present one illustration,
for $N_c=5$. This is a fit to the data sets with the smallest 15 $\xi$ values. Fig.~\ref{fig:priorsu5}
shows the LO and NLO quantities for ten choices of priors, set identical for $l_{12}$, $k_F$, and $k_M$.
The values, from left to right, are $(P,\Delta P)=(0.0,1.0)$, (0.0,2.0), (0.0,3.0),
(1.0,1.0), (1.0,2.0), (1.0,3.0),
(2.0,1.0), (2.0,2.0), (2.0,3.0),
(3.0,3.0). The fitted values for  $l_{12}$, $k_F$, and $k_M$ are shown in Fig.~\ref{fig:prior2su5}.
In our NLO fits in Sec.~\ref{sec:results}, we will keep the (1.0,2.0) prior.

We remark in passing that one lattice artifact which we looked for but did not observe was an extra term
in the relation $m_{PS}^2 = 2B m_q$ due to nonzero lattice spacing. This is an extra term on the right hand side
of the first equation in the set of Eq.~\ref{eq:realxfit}. We checked this most thoroughly for NNLO $x$ fits.
 The addition of  terms
$m_{PS}^2 = 2B m_q(\dots) + C_a (a^2/t_0)$ or 
$m_{PS}^2 = 2B m_q(\dots) + C_a (a/\sqrt{t_0})$ make a tiny change to the $\chi^2$ (comparing the same fit ranges)
compared to leaving them out. All that happens is that the uncertainties on the
other fit parameters (mostly $F$) grow and the central values drift by a $\sigma$ or so.
A broad prior on $C_a$ is needed to stabilize the fit.
The authors of Ref.~\cite{Budapest-Marseille-Wuppertal:2013vij} use an action with smeared gauge links
which is similar to ours and do not include this term either.

\section{Results from fits to  chiral perturbation theory\label{sec:results}}

\subsection{$SU(2)$ chiral perturbation theory\label{sec:resultsSU2}}

Our extraction of LEC's from our data sets is done by performing a series of fits varying the range of
quark masses, typically varying the maximum value of $\xi$ in the data set and using model averaging to
produce a set of results. The lattice data  for $f_{PS}$ and $m_{PS}^2$ are typically
very smooth functions of the quark mass. The fitting functions are also very smooth. It is usually
possible to get a good quality fit for any $\xi$ range. The issue is then whether the fit parameters
are stable under variation of fit ranges.

The values of $SU(N_f)$ LEC's were already a closed subject before we began our project.
They are  summarized by FLAG  \cite{FlavourLatticeAveragingGroup:2019iem}
 and come from data sets of much higher quality than ours. The study of $N_c>3$ is our goal.
We therefore use $N_c=3$ as a fiducial: can we reproduce FLAG results?

Recall that we are scaling all quantities with the appropriate power of $t_0$ to produce dimensionless quantities.
The 2019 FLAG review
tells us that $F=87$ MeV (in the ``93 MeV'' convention) so with $\sqrt{t_0}=0.15$ fm,
$\hat F = \sqrt{2} \times 87$ MeV $\times 0.15$ fm $\sim 0.09$ in our convention. The
condensate $\Sigma=F^2 B$ in the same convention.
The review
 quotes $\Sigma^{1/3}\sim 266(10)$ MeV (in $\overline{MS}$ at $\mu=2$ GeV)
and with $\hat F=0.09$, $\hat B \sim 2.05$. The review lists $l_3= 3.41(82)$
and $l_4 =4.40(28)$.

\begin{figure}
\begin{center}
\includegraphics[width=0.9\textwidth,clip]{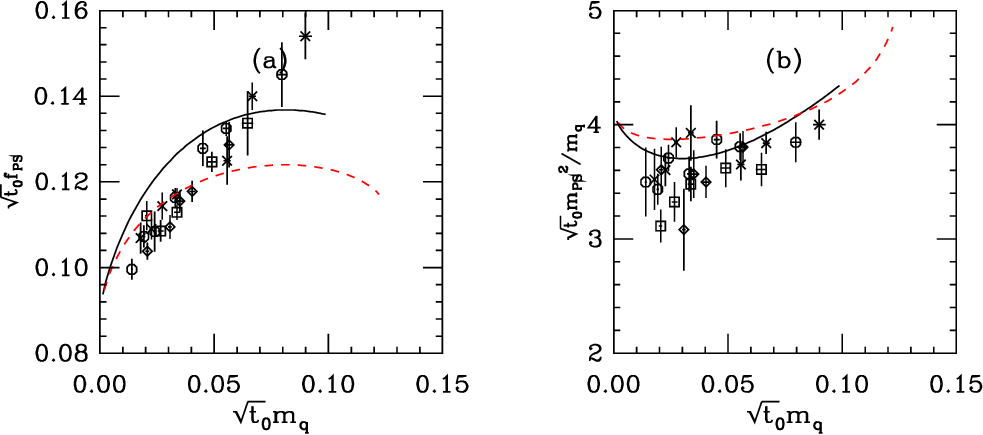}
\end{center}
\caption{$N_c=3$ data compared to NLO formulas
from the $x$-parameterization in the black solid lines black and the $\xi$ paramaeterization in red dashes
 with ``known'' (FLAG)
\protect{\cite{FlavourLatticeAveragingGroup:2019iem}} parameters for the LEC's.
(a) $\sqrt{t_0}f_{PS}$ versus $\sqrt{t_0}m_q$;
(b) $\sqrt{t_0}m_{PS}^2/m_q$ versus $\sqrt{t_0}m_q$. Symbols show
$\beta=5.25$ in squares,
$\beta=5.3$ in diamonds,
$\beta=5.35$ in octagons,
$\beta=5.4$ in crosses.
\label{fig:both}}
\end{figure}

 Before doing any fits, we can just compare our data to
NLO curves with these values for the LEC's. Fig.~\ref{fig:both} shows two plots:
$\sqrt{t_0}f_{PS}$ versus $\sqrt{t_0}m_q $ in panel (a)
and
$\sqrt{t_0}m_{PS}^2/m_q$ versus $\sqrt{t_0}m_q$ in panel (b).
The different plotting symbols correspond to different
values of the bare gauge coupling.

 The figure shows  one loop results with the FLAG
parameters for the LEC's. The black curves
show the NLO $x$ curve while the dashed red curves the  NLO $\xi$ curve. The data, especially
for $F_{PS}$, are much straighter than either NLO fit would favor. In an NLO fit to the data, 
the fit parameter for $l_4$ will drift upward as higher mass points are kept, to try to straighten the curve.
The data sets we collected mostly lie outside
the range where NLO chiral perturbation theory is applicable. To fit all our data sets
it is necessary to do NNLO fits.

\begin{figure}
\begin{center}
\includegraphics[width=0.8\textwidth,clip]{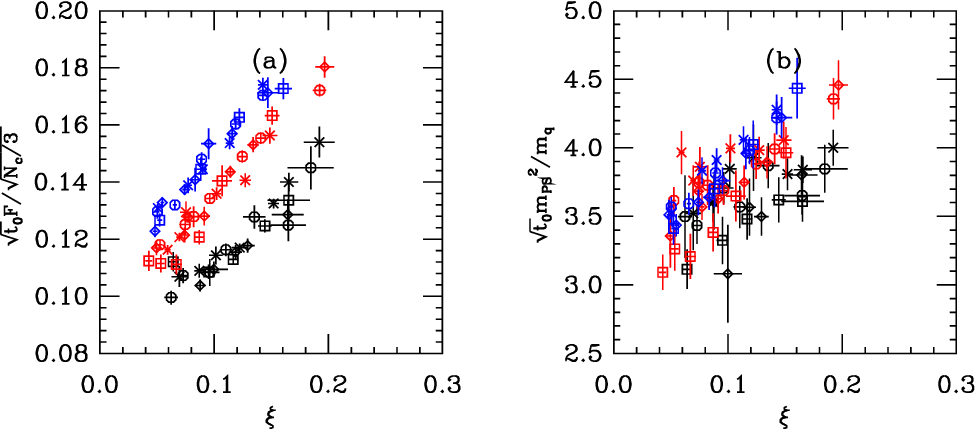}
\end{center}
\caption{Plots of (a) $\sqrt{t_0}f_{PS}/\sqrt{N_c/3}$ and  (b) $\sqrt{t_0}m_{PS}^2/m_q$ versus $\xi=m_{PS}^2/(8\pi^2 f_{PS}^2)$.
Colors label different $N_c$ values (black for $N_c=3$, red for $N_c=4$ and blue for $N_c=5$),
and the different symbols label the different bare gauge couplings as described in the text.
\label{fig:fmpi2mqxi}}
\end{figure}

As a second preliminary picture we display $\sqrt{t_0}f_{PS}/\sqrt{N_c/3}$ 
and $\sqrt{t_0}m_{PS}^2/m_q$ versus $\xi$,
in Fig.~\ref{fig:fmpi2mqxi}. 
The quantities plotted on the $x$ and $y$ axes of these plots are quite degenerate,
but the figure does display the landscape of the data.
The different plotting symbols label the different beta values (and hence different
 lattice spacings) of the data sets.
The catalog is:
\begin{itemize}
\item
For $N_c=3$, shown in black,
squares for $\beta=5.25$, diamonds show $\beta=5.3$, octagons show $\beta=5.35$,
crosses show $\beta=5.4$.
\item
For $N_c=4$, shown in red, squares are  for $\beta=10.0$, diamonds for $\beta=10.1$, octagons for $\beta=10.2$,
crosses for $\beta=10.3$.
\item
For $N_c=5$, shown in blue, squares arecfor $\beta=16.2$, diamonds for $\beta=16.3$, octagons for $\beta=16.4$,
crosses for $\beta=16.6$.
\end{itemize}

\begin{figure}
\begin{center}
\includegraphics[width=0.8\textwidth,clip]{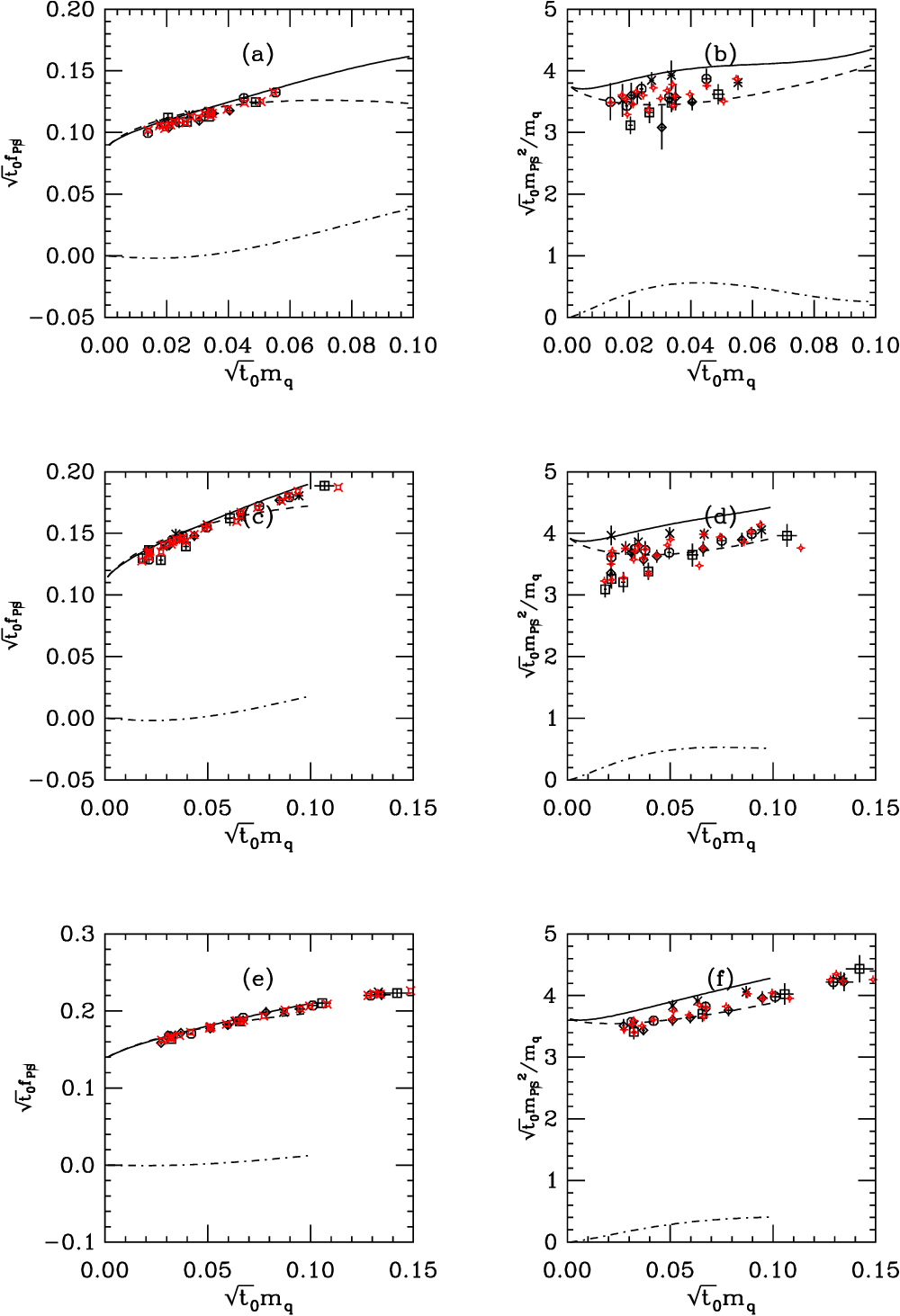}
\end{center}
\caption{NNLO $x$ fits showing  $\sqrt{t_0}f_{PS}$ and $\sqrt{t_0} m_{PS}^2/m_q$   vs $\sqrt{t_0}m_q$.
(a)  and (b) show $N_c=3$, (c) and (d) show  $N_c=4$, and (e) and (f) show  $N_c=5$ data sets. 
The fits involve the seven parameters of the
 NNLO expression, with the NNLO LEC's stabilized by priors as described in the text and two $O(a^2)$ correction terms as
described in Eq.~\ref{eq:realxfit}. 
The different plotting symbols label the different beta values of the data sets and are given in the text.
Red points are the fitted values associated with the black data points. 
The lines show the continuum limit of the fitting functions. The
decomposition of the $SU(N_f)$ fitting functions for $f_{PS}$ and $m_{PS}^2/m_q$ (shown as solid lines) is split
into their NLO (dashed lines) and NNLO (dash-dotted lines) components.
\label{fig:fitnnlox}}
\end{figure}

 We begin with a set of typical results for NNLO $x$ fits, Fig.~\ref{fig:fitnnlox}.
The data in these fits has been selected to be less than some chosen value of $\xi_{max}$.
The fits involve the seven parameters of the
 NNLO expression, with the NNLO LEC's stabilized by priors as described in the text, and
 two $O(a^2)$ correction terms as
given in Eq.~\ref{eq:realxfit}. The two sets of points show
$\sqrt{t_0} m_{PS}^2/m_q$  and $\sqrt{t_0}f_{PS}$ vs $\sqrt{t_0}m_q$.
The different plotting symbols label the different beta values (and hence different
 lattice spacings) of the data sets as we have listed.
The red points are the fit values associated with each data point. The solid lines 
are the continuum result. The difference between the line
and the data points is due to the nuisance parameters $C_F$ and $C_B$ in Eq.~\ref{eq:realxfit}.

\begin{figure}
\begin{center}
\includegraphics[width=0.8\textwidth,clip]{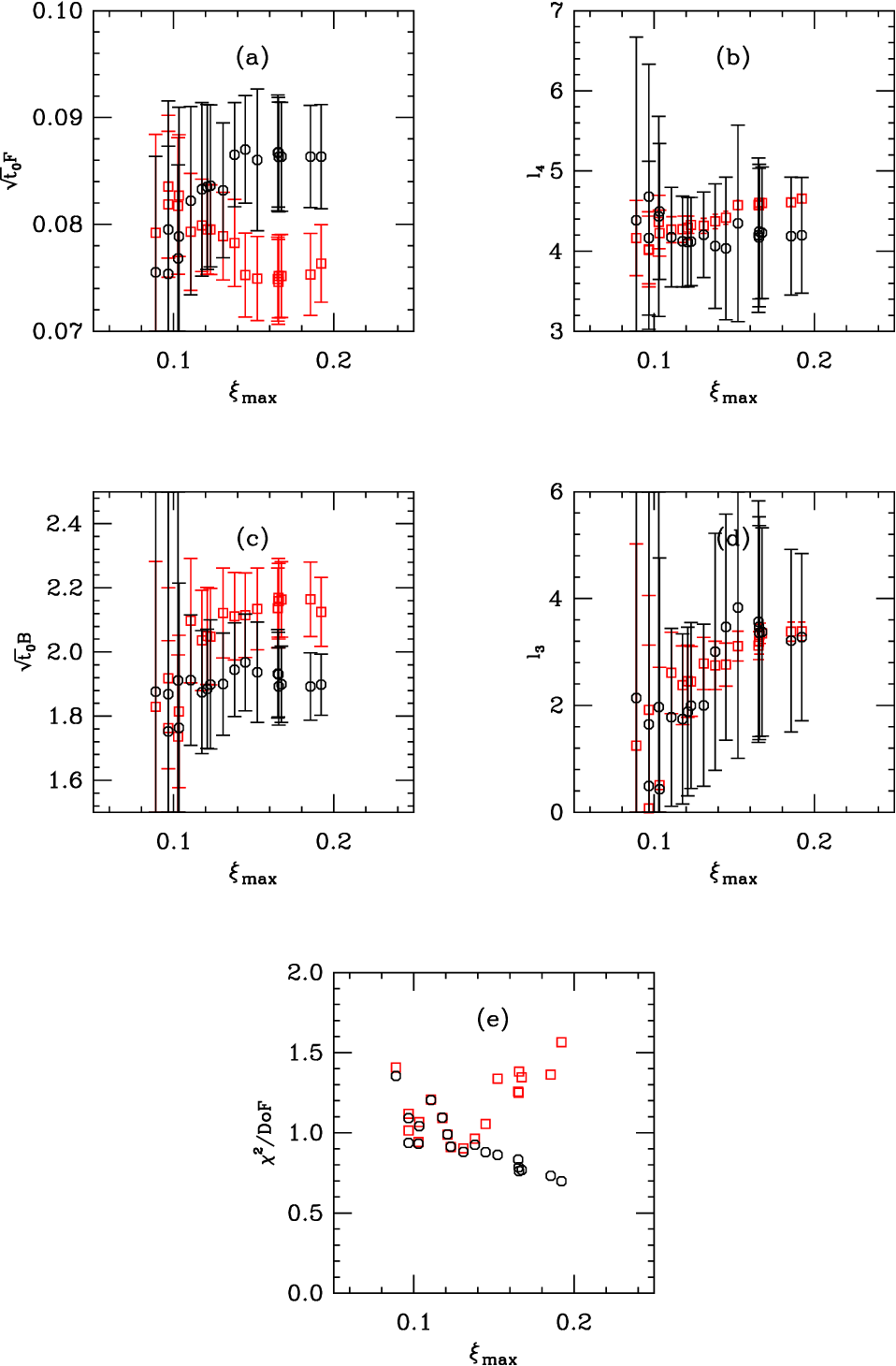}
\end{center}
\caption{Results of NLO and NNLO $SU(2)$ $x$ fits to $N_c=3$  data sets versus $\xi_{max}$.
The red squares show NLO fits and the black octagons show NNLO fit results.
(a) $\sqrt{t_0}F$,
(b) $l_4$,
(c) $\sqrt{t_0}B$,
(d) $l_3$,
(e) the $\chi^2$ per degree of freedom.
\label{fig:rangesu3x}}
\end{figure}

To determine the LEC's, we organized a set of NNLO fits  by sorting each $N_c$'s
data into a file with increasing $\xi$ and performed a series of fits beginning with a subset of points with the
smallest $\xi$ values and extending up to some maximum $\xi$. 
We monitored the $\chi^2$ per degree of freedom and confidence level
of the fits. Fig.~\ref{fig:rangesu3x} shows results from $N_c=3$ for $B$, $F$, $l_3$ and $l_4$, along with the
chi-squared per degree of freedom.
  The flatness of the fit quantities versus $\xi_{max}$ indicates that we can
perform model averaging over our suite of fits to determine the LEC's.

\begin{figure}
\begin{center}
\includegraphics[width=0.8\textwidth,clip]{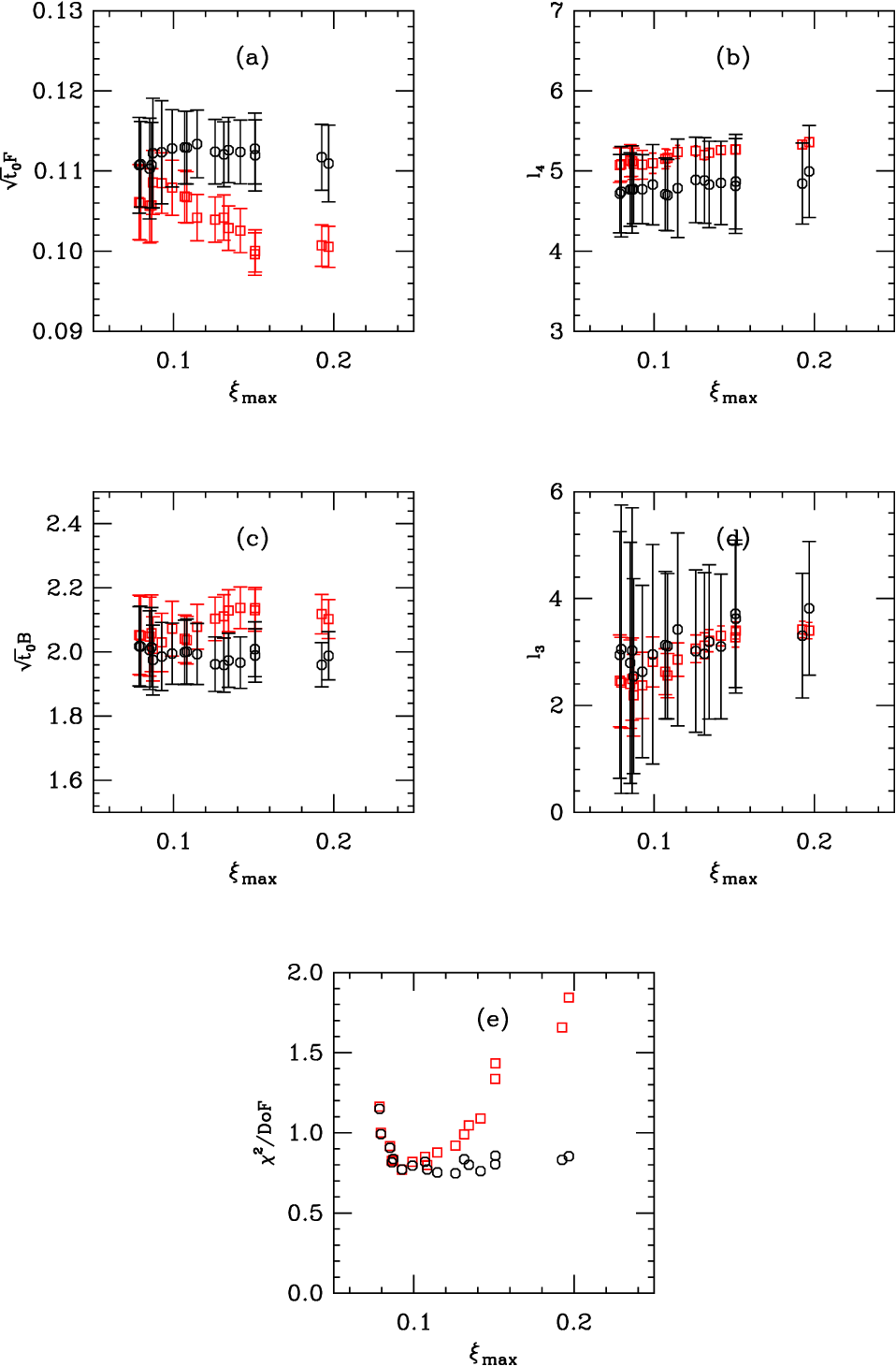}
\end{center}
\caption{Results of  NNLO $x$  fits to $SU(4)$ data sets versus $\xi_{max}$, as in Fig.~\protect{\ref{fig:rangesu3x}}.
\label{fig:rangesu4x}}
\end{figure}

\begin{figure}
\begin{center}
\includegraphics[width=0.8\textwidth,clip]{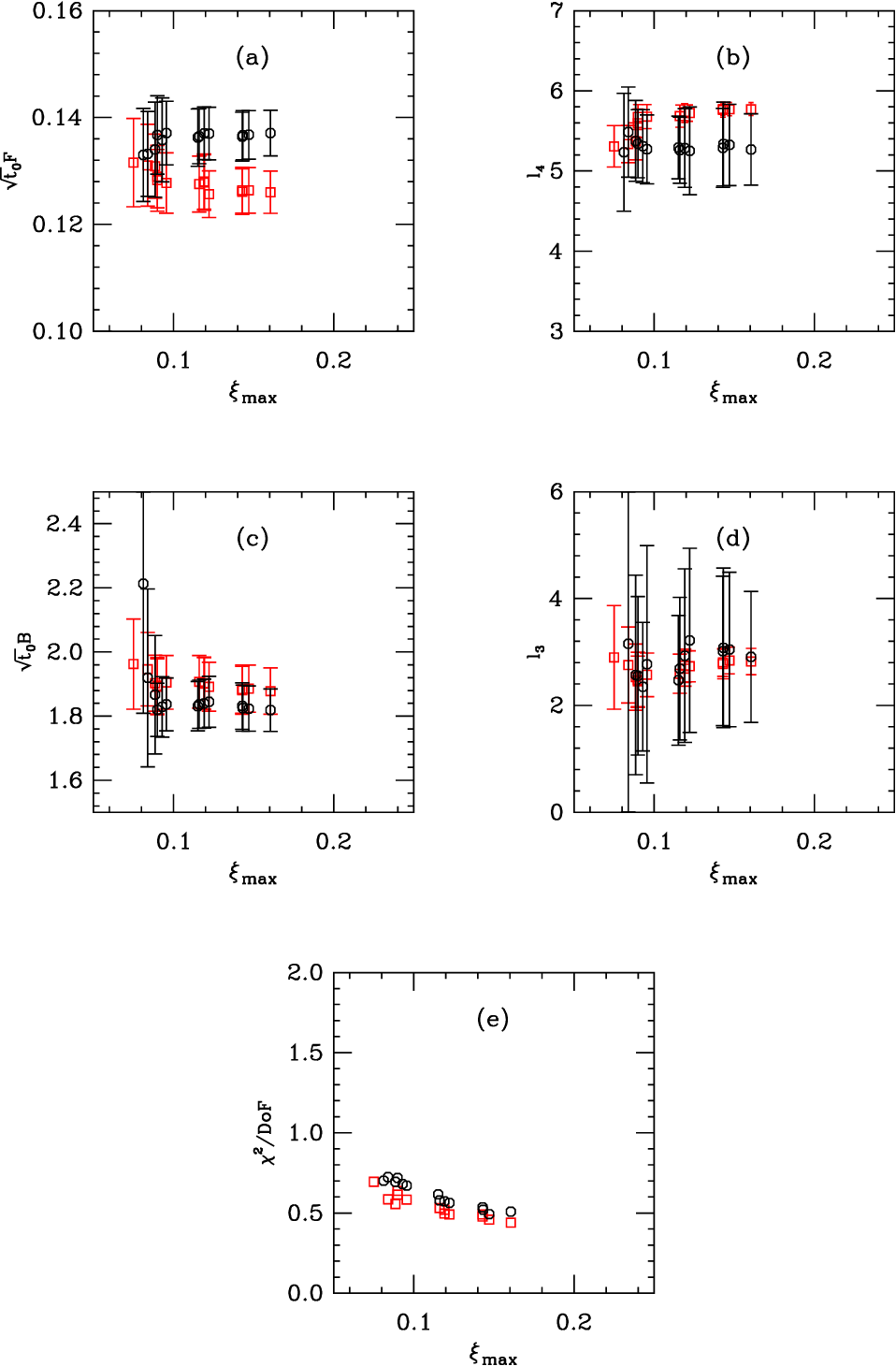}
\end{center}
\caption{Results of  NNLO $x$ fits to $SU(5)$ data sets versus $\xi_{max}$, as in Fig.~\protect{\ref{fig:rangesu3x}}.
\label{fig:rangesu5x}}
\end{figure}

\begin{figure}
\begin{center}
\includegraphics[width=0.8\textwidth,clip]{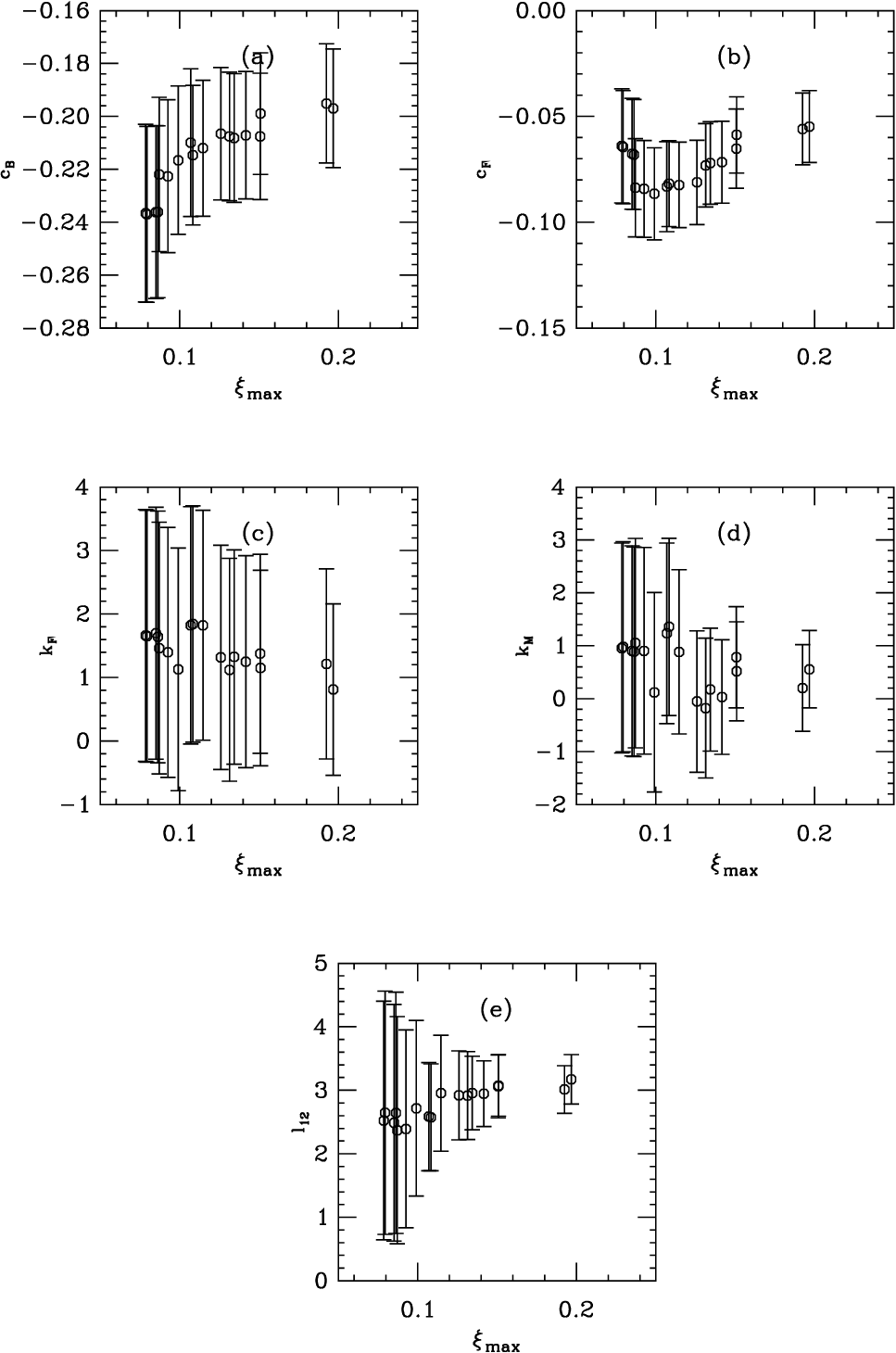}
\end{center}
\caption{More results of NNLO $x$ fits to $SU(4)$ data sets versus $\xi_{max}$: 
(a) $c_B$,  the $a^2$ correction to $B$
(b) $c_F$, the $a^2$ correction to $F$,
and the three NNLO LEC's (controlled by priors in the fits)
(c) $k_F$,
(d) $k_M$,
(e) $l_{12}$.
\label{fig:range2su4x}}
\end{figure}

Figs.~\ref{fig:rangesu4x} and \ref{fig:rangesu5x} show the same information, but for $N_c=4$ and 5.
We show one representative example of our determination of the cutoff dependent
terms $c_B$ and $c_F$ and the NNLO parameters $k_F$, $k_M$ and $l_{12}$, for $N_c=4$, in Fig.~\ref{fig:range2su4x}.
Recall that the NNLO fitted parameters are strongly controlled by priors.

\begin{figure}
\begin{center}
\includegraphics[width=0.9\textwidth,clip]{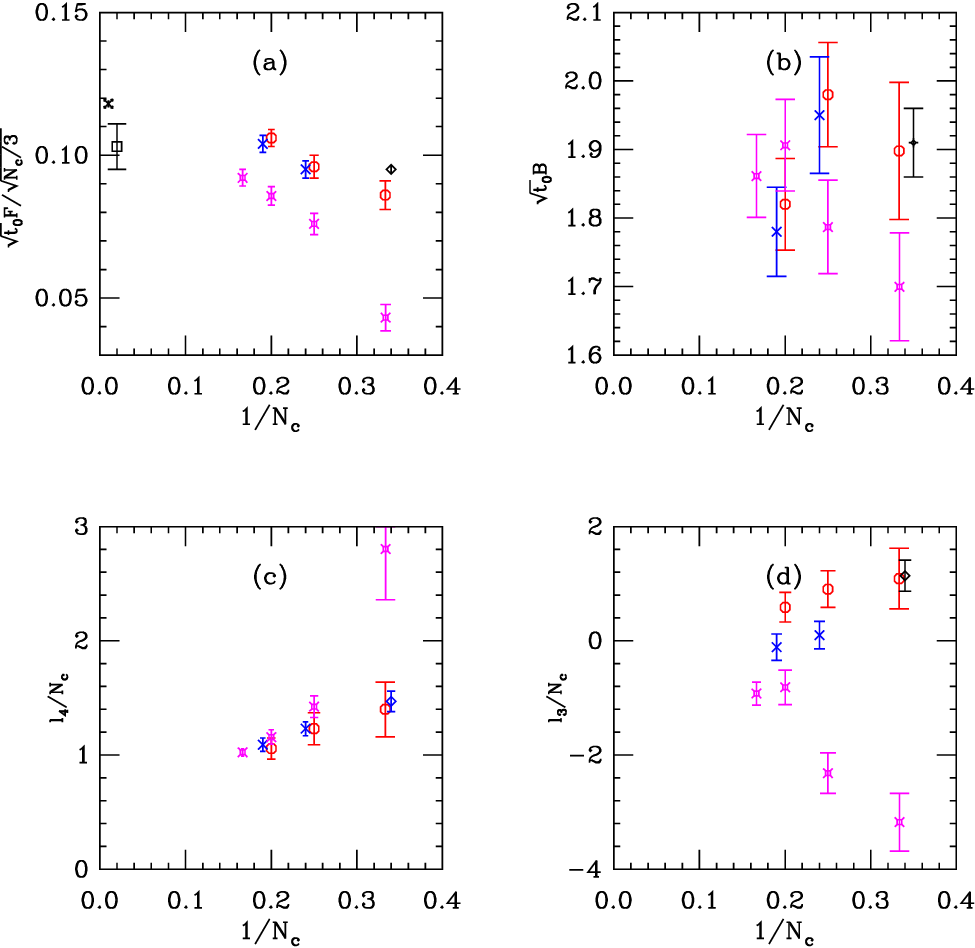}
\end{center}
\caption{  LEC's from fits of NNLO $SU(2)$ chiral perturbation theory
 for $N_c=3$, 4, and 5,
along with other relevant results for comparison.
(a) $\sqrt{t_0}F/\sqrt{N_c/3}$,
(b) $\sqrt{t_0}B$,
(c) $l_4/N_c$,
(d) $l_3/N_c$.
The plotting symbols are
red octagons, for our $SU(2)$ results;
blue crosses, our $U(2)$ results, converted to $SU(2)$ LEC's.
Other relevant data are also shown (in black). High precision $N_f=2$, $N_c=3$ results are
diamonds \protect{\cite{FlavourLatticeAveragingGroup:2019iem}},
and
fancy diamonds, \protect{\cite{Budapest-Marseille-Wuppertal:2013vij}}.
There are
two large $N_c$ limits of quenched data, the large volume results of
\protect{\cite{Bali:2013kia}} as squares
and the small volume ones of
\protect{\cite{Perez:2020vbn}} as fancy crosses.
Purple fancy squares  show $N_f=4$ results from \protect{\cite{Hernandez:2019qed}}.
\label{fig:final}}
\end{figure}

Results are listed in listed in Table \ref{tab:xlec} and displayed in Fig.~\ref{fig:final}.

The uncertainties on $l_3$ are much larger than those of $l_4$. 
This is similar to what is seen in the FLAG averages for $N_c=3$ \cite{FlavourLatticeAveragingGroup:2019iem}.


As a final check of the consistency of our results, we break apart the 
NNLO predictions for $f_{PS}$ and $m_{PS}^2/m_q$
into their LO+NLO and NNLO  components and plot them. This is shown in 
Fig.~\ref{fig:fitnnlox}. The NNLO piece
remains small over the range of $\sqrt{t_0}m_q$ values used in fits.

\subsection{$U(2)$ chiral perturbation theory\label{sec:resultsU2}}

The analysis path for $U(2)$ chiral fits exactly parallels that for $SU(2)$. We performed fits to the NLO
and NNLO formulas, studying individual $N_c$ data sets. We included additional nuisance parameters to
account for finite lattice spacing effects. Specifically,
\beea
m_{PS}^2 &=& (1+c_B\frac{a^2}{t_0})( M^2_{NLO} + M^2_{NNLO}) \nonumber \\
f_{PS} &=&  (1+c_F\frac{a^2}{t_0})  (F_{NLO} + F_{NNLO}) \nonumber \\
\label{eq:ualll}
\eea
where the NLO and NNLO terms
are given in Eqs.~\ref{eq:unlo}-\ref{eq:unnlo}.
 The NLO fits involve six parameters $B$, $F$, $l_3^{(0)}$,
$l_4^{(0)}$, $c_B$ and $c_F)$. The NNLO fits add two more: $T_F$ and $T_M$,   
 for eight parameters, and replaces the
$l_i^{0)}$'s by $l_i^{(0)}+l_i^{(1)}/N_c$.

 We need one more input parameter, the quenched topological susceptibility $\chi_T$. 
Ref.~\cite{Ce:2016uwy} measured it to be 
\bee
t_0^2 \chi_T= 7\times 10^{-4}.
\ee
 (See also \cite{Ce:2015qha} for earlier determinations of $\chi_T$.)

We experimented with priors for  $T_F$ and $T_M$. We discovered that if the fitting range in $\xi$ or 
$\sqrt{t_0}m_q$ was large, no priors were needed, while when the range became small, setting
a prior $0.2 \pm 0.2$ for $T_F$ and $-0.2\pm 0.2$ for $T_M$ stabilized the fit.

Notice that the NLO fits are ``fits to a straight line''  (plus lattice artifacts)  of our data.

As in the case of the $SU(2)$, we can generally find fits which have low chi-squares
 to any data set; the issue is whether the fit parameters are stable under data set size.
A second consideration is whether the fit makes sense in that the NNLO contribution is small
compare to the NLO one.

\begin{figure}
\begin{center}
\includegraphics[width=0.8\textwidth,clip]{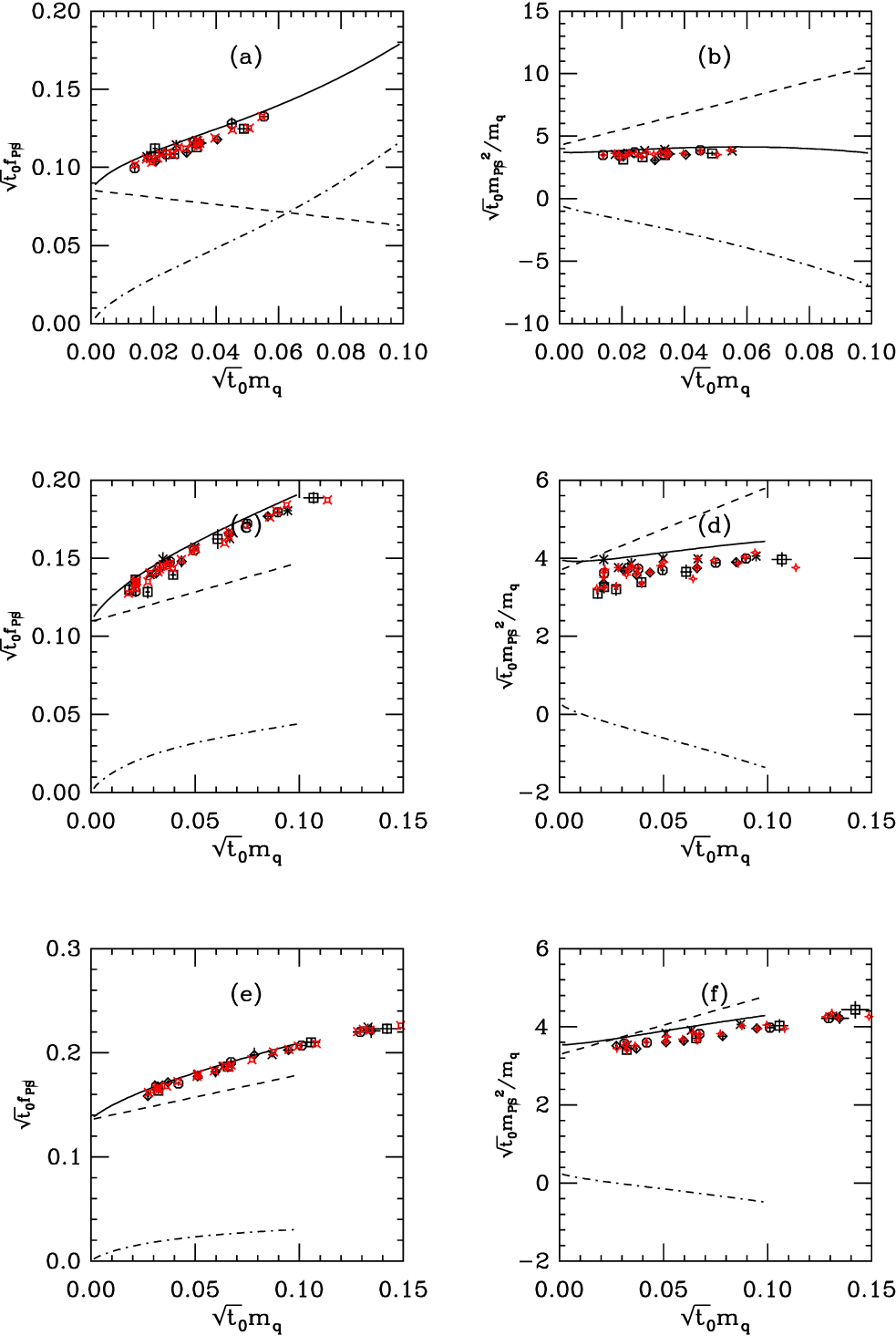}
\end{center}
\caption{Decomposition of the $U(N_f)$ fitting functions for $f_{PS}$ and $m_{PS}^2/m_q$ (shown as solid lines)
into their NLO (dashed lines) and NNLO (dash-dotted lines) components. 
(a) and (b) $N_c=3$;
(c) and (d) $N_c=4$;
(e) and (f) $N_c=5$.
The data are shown with the same plotting symbols as in Fig.~\protect{\ref{fig:fitnnlox}}.
\label{fig:ufits}}
\end{figure}

$N_c=3$ is a special case: $U(2)$ chiral fits fail. This can be seen in Fig.~\ref{fig:ufits}, 
where we break up a typical
NNLO fit into its component parts. For $N_c=3$ the actual NNLO component is huge compared to the NLO piece.
The fitted $B$ also has a very large uncertainty as it tries to compensate for the incorrect functional form demanded by
the fitting function. We include the data with the same plotting symbols as in Fig.~\ref{fig:fitnnlox}.

This leaves us $N_c=4$ and 5. Fig.~\ref{fig:ufits} shows that the size of the
 various contributions is consistent with
a small NNLO correction and that the situation is more improved for $N_c=5$ than for $N_c=4$.

\begin{figure}
\begin{center}
\includegraphics[width=0.8\textwidth,clip]{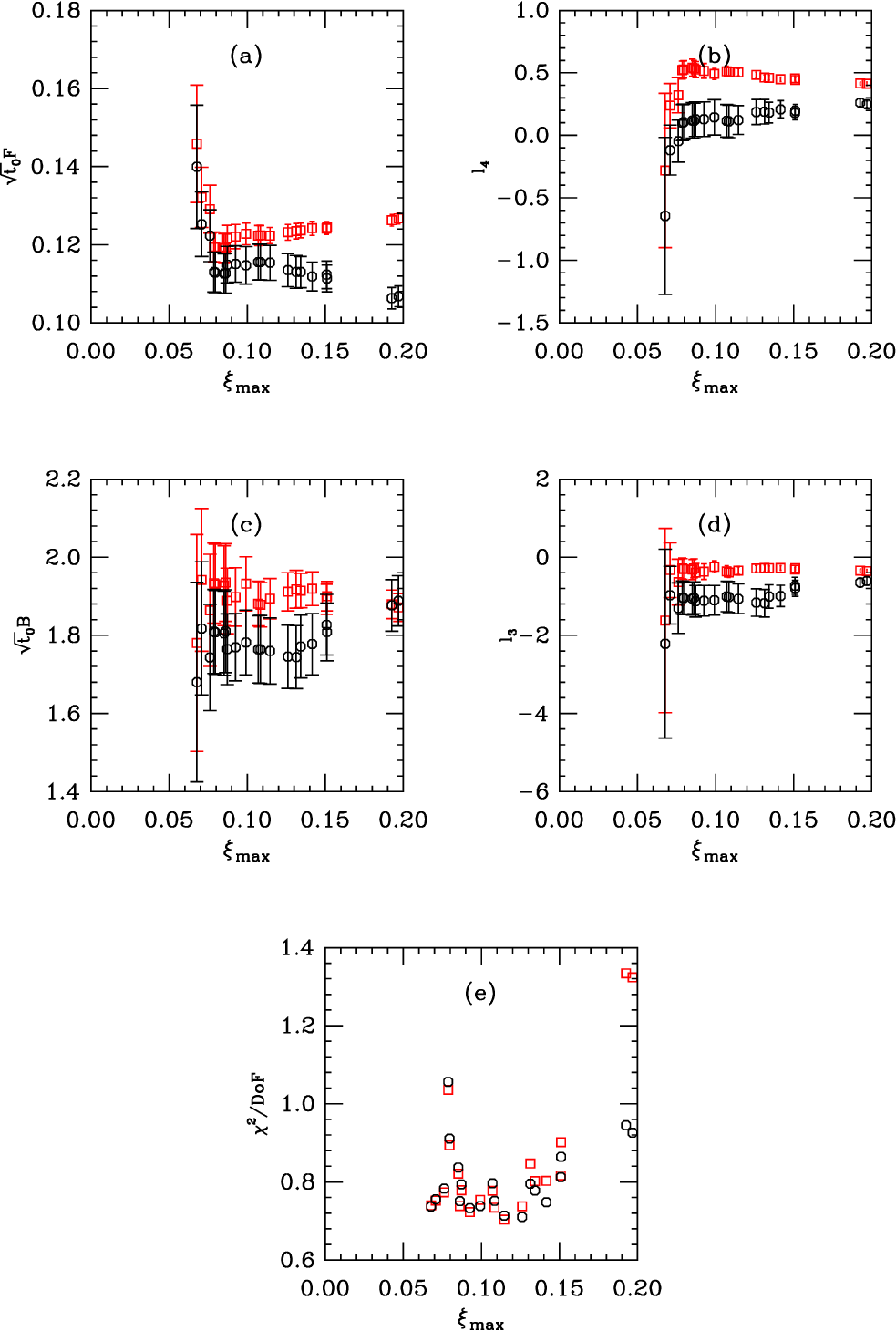}
\end{center}
\caption{Results of NLO (red squares) and NNLO (black octagons)  $U(2)$ fits to $N_c=4$ data sets versus $\xi_{max}$.
(a) $\sqrt{t_0}F$,
(b) $l_4^{(0)}$ or $l_4^{(0)}+l_4^{(1)}/N_c$,
(c) $\sqrt{t_0}B$,
(d) $l_3^{(0)}$ or $l_4^{(0)}+l_3^{(1)}/N_c$,
(d) $l_3$,
(e) the $\chi^2$ per degree of freedom.
\label{fig:rangesu4u}}
\end{figure}

\begin{figure}
\begin{center}
\includegraphics[width=0.8\textwidth,clip]{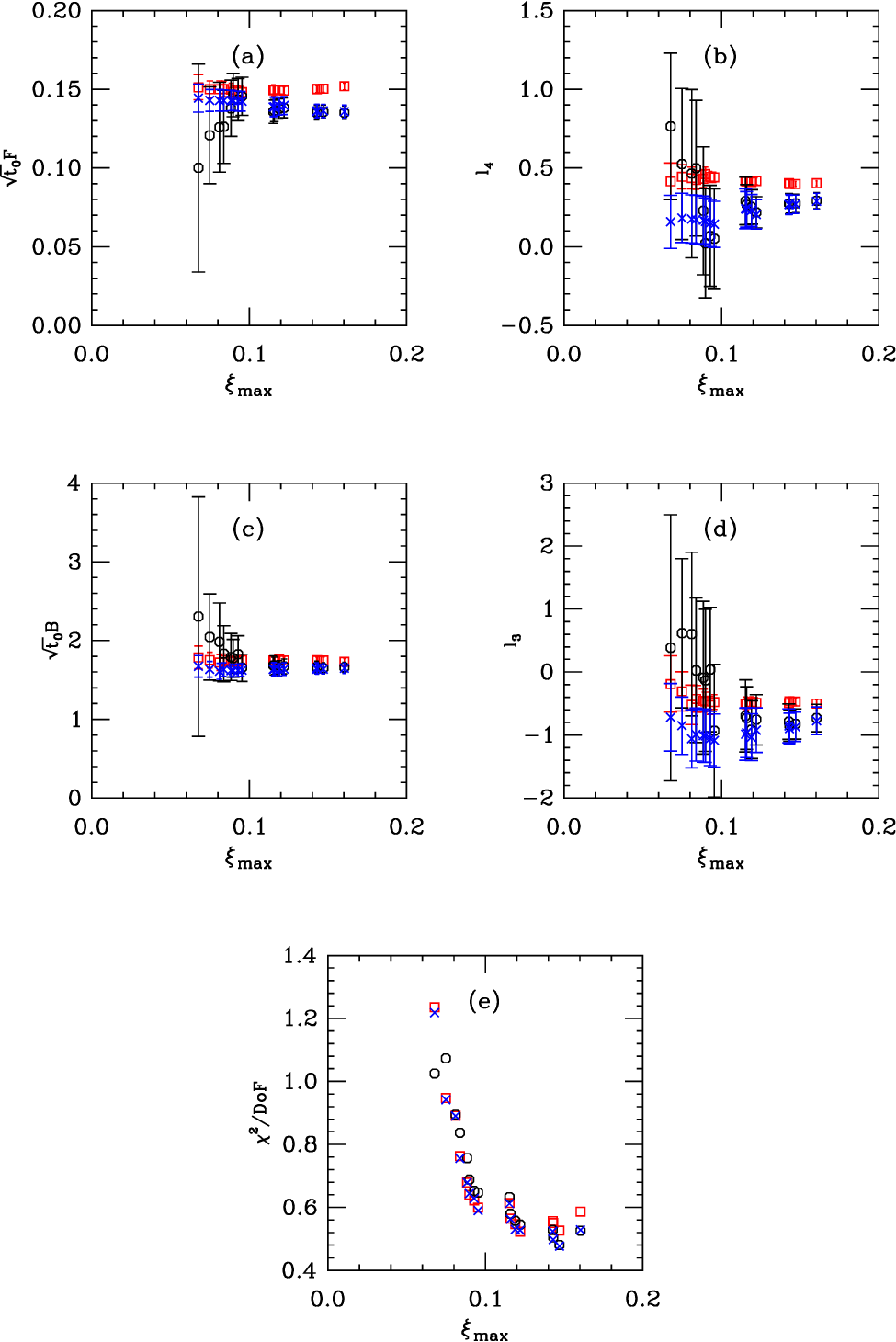}
\end{center}
\caption{Results of NLO and NNLO  $U(2)$ fits to $N_c=5$ data sets versus $\xi_{max}$. 
The red squares are NLO fits, the black octagons are NNLO fits without priors, and
the blue crosses are NNLO fits with priors.
(a) $\sqrt{t_0}F$,
(b) $l_4^{(0)}$ or $l_4^{(0)}+l_4^{(1)}/N_c$,
(c) $\sqrt{t_0}B$,
(d) $l_3^{(0)}$ or $l_4^{(0)}+l_3^{(1)}/N_c$,
(d) $l_3$,
(e) the $\chi^2$ per degree of freedom.
\label{fig:rangesu5u}}
\end{figure}

We proceed to results. Figs~\ref{fig:rangesu4u}-\ref{fig:rangesu5u} show results 
of individual NLO and NNLO fits out to
values of $\xi_{max}$. The NLO fits drift when $\xi_{max}$ becomes greater than about 0.1.
These fits will be model-averaged once a range is chosen. We then compare 
NLO and NNLO fits at $\xi_{max}=0.107$ with NNLO fits at $\xi_{max}=0.20$ ($N_c=4)$ or 
NLO and NNLO fits at $\xi_{max}=0.09$ with NNLO fits at $\xi_{max}=0.16$ ($N_c=5$). Fitted values for the LEC's
agree at the one standard deviation level.
Results from the NNLO fits at large $\xi_{max}$ are listed in Table \ref{tab:ulec}.

One final thing we must do is connect the $U(2)$ LEC's to the $SU(2)$ ones. The $F$ values need no
conversion. The other LEC's need a shift and rescaling according to Eq.~\ref{eq:unsus} and \ref{eq:sushift}.
With the dimensionless $U(2)$ scheme point $\mu= t_0 (8\pi^2 F^2)=0.64$ and the $SU(2)$ $\mu= t_0m_{\pi}(phys)=0.011$,
the converted expressions are recorded in Table~\ref{tab:ulec}. As we saw in the $SU(2)$ case, the
results for $l_3$ are quite a bit noisier than the ones for $l_4$.

\subsection{Discussion of results \label{sec:discuss}}
 We collect our results in a panel of figures, Fig.~\ref{fig:final}.
Our results are presented in $\overline{MS}$ at a scale of 2 GeV (equivalent, for $N_c>3$, of course)
and we have converted the $l_i$'s to a scale $\mu^2=m_\pi^2$.
The $U(2)$ parameters have been matched to $SU(2)$ ones, so the plots show $SU(2)$ LEC's.

We have included data from other sources. At $N_c=3$ these are the FLAG
\cite{FlavourLatticeAveragingGroup:2019iem} numbers. They do not quote a direct average for $B$ so
 we substitute the measurement of $B$ 
from BMW \cite{Budapest-Marseille-Wuppertal:2013vij} as comparison.

We also added two $N_c \rightarrow \infty$ results, from the large $N_c$-small volume simulations of
Ref.~\cite{Perez:2020vbn} and from the large volume quenched simulations of Ref.~\cite{Bali:2013kia}.

The authors of Ref.~\cite{Perez:2020vbn} performed simulations in the context of the 
twisted Eguchi-Kawai matrix model at
$N_c=169$, 289, and 361 at several lattice spacings per $N_c$. They
 give the large $N_c$ limiting results
$\sqrt{8 t_0 \sigma}=1.078(9)$ and $f_{PS}/\sqrt{\sigma}\sqrt{3/N_c}=0.22(1)(2)$
where $\sigma$ is the string tension and the two uncertainties are statistical and systematic.
They use the ``93 MeV'' definition for the decay constant, so we convert their numbers to
\bee
\sqrt{\frac{3}{N_c}} \sqrt{t_0}F =
 \sqrt{2}[\frac{F}{\sqrt{\sigma}} \frac{\sqrt{8 t_0 \sigma}}{\sqrt{8}} ] =
0.118(1)
\ee
(combining errors in quadrature for the plot).

Ref.~\cite{Bali:2013kia} has quenched data for $N_c=2-7$ and 17, at one lattice spacing, $a\sim 0.1$ fm.
They quote a large $N_c$ limit of
\bee
\frac{1}{\sqrt{2}}\sqrt{\frac{3}{N_c}} \frac{F}{\sqrt{\sigma}}= 0.2174(30)
\ee
with an additional eight per cent error on the string tension, which is the dominant uncertainty.
 With a choice
of $\sqrt{\sigma}=440$ MeV and $t_0=0.15$ fm, this is the limiting value 
\bee
\lim_{N_c\rightarrow\infty} \sqrt{\frac{3}{N_c}} \sqrt{t_0}F=0.103(8).
\ee
We note that the authors of  Ref.~\cite{Bali:2013kia} extrapolate all their data in $1/N_c^2$.

Next, we attempt a comparison with the results of Hernandez et al 
\cite{Hernandez:2019qed}. Their $N_f=4$ data sets are fit to $U(4)$ chiral 
perturbation theory and they perform interpolations in $N_f$ assuming that
$N_f$ appears in the combination $N_f/N_c$ (so that tuning $N_c$ is a proxy
for varying $N_f$).

They have tables of $aF/\sqrt{N_c}$ which we convert to $\sqrt{t_0}\sqrt{3/N_c}F$
in our conventions using $\sqrt{t_0}=0.15$ fm and their $a=0.075$ fm.
Their $N_f=4$ results are lower than our $N_f=2$ ones. The coefficient
 of $N_f/N_c$ in their fitting function is negative
and their $N_f=2$ extrapolations are 0.093(3) fo $N_c=3$,
0.101(3) for $N_c=4$ and $0.105(3)$ for $N_c=5$. Their extrapolation to infinite $N_c$
is 0.124(3) - 0.130(3) for the two functional forms they use.

We present $\sqrt{t_0}B = Z_s \sqrt{t_0}/a (aB)$ from their tables, converted to
the $SU(N_f)$ $B$, in Fig.~\ref{fig:final}b. There seems to be little $N_f$ dependence on this 
quantity -- which their $N_f/N_c$ parameterizations also show.

We translated their data for $L_F$ into $l_4(\mu=m_\pi^2)$ by
\bee
l_4(N_f=4,\mu) = \log\frac{8\pi^2F^2}{m_\pi^2}+ \frac{64\pi^2}{(N_f/2=2)} L_F.
\ee
Fig.~\ref{fig:final}c shows that their $N_f=4$ results for $N_c >3$
 are consistent with ours. Their prediction of $N_c=3$, $N_f=2$ numbers
($l_4=5.1(3)$ or 4.1(11) for two fits) are consistent with ours.

Finally, $l_3$. We again convert their $U(4)$ quantity to an $SU(4)$ one and plot it in 
panel d of Fig.~\ref{fig:final}. This quantity is apparently strongly $N_f$ dependent,
in addition to being very noisy.
They quote an $N_c=3$, $N_f=2$ value of $l_4=0.4\pm 1.6$ (at $\mu=m_\pi$)
to be compared with our $3.2\pm 1.5$ and FLAG's 3.41(82).

 It's a bit dangerous to extrapolate our data to $N_c\rightarrow \infty$, we think.
But (just to save  readers the work of doing it themselves)
a linear fit to the $SU(2)$ values for $\sqrt{t_0}B$ and $\sqrt{t_0}\sqrt{3/N_c}F$
of the form $c_1+c_2/N_c$ gives large $N_c$ values of $\sqrt{t_0}B=1.72(22)$
and   $\sqrt{t_0}\sqrt{3/N_c}F=0.136(10)$.
A glance at Fig.~\ref{fig:final} shows that these predictions seem to be unexpected.
Of course, a convincing plot would benefit from 
continuum predictions from all the possible
approaches (listed at the beginning of the paper) to the large $N_c$ limit.
Presumably all approaches would converge to the same large $N_c$ limit, but would do so
in different ways, which might illustrate how varying numbers of fermion flavors
affect the LEC's.

\section{Conclusions\label{sec:conclude}}

The results shown in Fig.~\ref{fig:final} 
show that we have not answered the list of questions which motivated this project,
 though we may have made a start:
\begin{enumerate}
\item How do the low energy constants of the chiral effective theory scale with $N_c$? What is their limit?

The expected leading scaling $B\sim N_c^0$, $F\sim \sqrt{N_c}$, $l_3,$ $l_4\sim N_c$ seem to hold.
 We can see subleading
$N_c$ behavior in $F$, $l_3$, and $l_4$. With three values of $N_c$ we cannot say anything about
corrections beyond $1/N_c$.
\item Is there a crossover to $U(2)$ chiral behavior? 

This we cannot tell.  $U(2)$ fits fail for $N_c=3$.
We see that in our $N_c=4$ and 5 data sets both chiral expansions give comparable results for the LEC's,
when they are converted to a common scheme. We are aware of no direct calculations of the mass
of the flavor singlet meson for $N_c>3$, so the $U(2)$ fitting functions depend on
 the validity of the Witten - Veneziano
relation.
\end{enumerate}

What could we have done better (and why)? The list is obvious:
smaller fermion masses would have been a big help, to be able to do chiral fits deeper in the NLO regime.
This then implies a need for bigger lattice volumes, so as not to be compromised by finite size effects.
Fortunately, the larger the value of $N_c$, the less this is an issue.
Our large lattice spacings led to big uncertainties in the value of lattice to continuum 
matching factors, and the $SU(3)$ literature makes the obvious point that smaller lattice spacing
allows more controlled matching at larger momentum scales.
And finally, more values of $N_c$ would certainly have been helpful. Trying to extrapolate three 
values of $N_c$ leaves little room to deal with the possibility that quantities scale
nonlinearly in $1/N_c$. We suspect, though, that data sets of the size of the ones reported in FLAG
for $SU(3)$ could complete the story of the chiral and continuum limit of large $N_c$ QCD.

Of course, to really give high precision answers to the questions we have asked will probably 
involve some kind of super - analysis involving a variety of approaches including
 simulations at several fixed $N_f$ values,
as well as quenched simulations in small and large volumes.
But in the meantime, we think that the low-statistics take away continues to 
be that $N_c=3$ QCD is not that different
from its large $N_c$ limit.

\clearpage

\begin{table}
\begin{tabular}{c c c c c c c c c c c c}
\hline
 $\kappa$ & $L$ & $am_q$  & $am_{PS}$ & $af_{PS}$ & $t_0/a^2$ &  $C_1$ & $C_2$ & $C_3$ & $Z_A$ & $Z_m$ & $N_{conf}$ \\
\hline
$\beta=5.25$ & & & & & & & & &  & \\
\hline
  0.1284 & 16 &  0.0628(5) &  0.4743(31) &  0.519(28)  &  0.927(3) &  0.423 & 0.196 & -0.120 & 0.984(11) &  1.011(37)  & 50   \\ 
   0.1288 & 16 &  0.0476(6) &  0.4135(34) &  0.4937(80)  &  0.981(3) &  0.508 & 0.438 & 0.314 & 0.962(8) &  1.010(40)  & 90   \\ 
   0.1292 & 16 &  0.0324(5) &  0.3371(27) &  0.4479(64)  &  1.045(3) &  0.564 & 0.417 & 0.106 & 0.955(7) &  1.024(37)  & 90   \\ 
   0.1294 & 16 &  0.0253(5) &  0.2926(36) &  0.4234(90)  &  1.076(4) &  0.546 & 0.567 & 0.280 & 0.967(8) &  1.033(42)  & 90   \\ 
   0.1296 & 16 &  0.0189(4) &  0.2488(36) &  0.443(13)  &  1.128(7) &  0.393 & 0.608 & 0.093 & 0.946(7) &  1.064(30)  & 90   \\ 
\hline
$\beta=5.3$ & & & & & & & & &  & \\
\hline
   0.1280 & 16 &  0.0470(4) &  0.3956(24) &  0.438(17)  &  1.249(10) &  0.034 & 0.466 & -0.073 & 0.978(10) &  1.026(34)  & 90   \\ 
   0.1284 & 16 &  0.0334(3) &  0.3213(36) &  0.3969(59)  &  1.326(5) &  0.251 & 0.583 & 0.216 & 0.980(14) &  1.033(31)  & 110   \\ 
   0.1285 & 16 &  0.0288(13) &  0.3022(31) &  0.3961(74)  &  1.381(5) &  0.454 & 0.407 & 0.211 & 0.961(9) &  1.038(30)  & 90   \\ 
   0.1286 & 16 &  0.0244(3) &  0.263(15) &  0.3705(87)  &  1.414(6) &  -0.006 & -0.112 & 0.065 & 0.966(8) &  1.071(24)  & 90   \\ 
   0.1288 & 24 &  0.0177(3) &  0.2333(44) &  0.3580(64)  &  1.449(5) &  0.310 & 0.457 & 0.113 & 0.959(6) &  0.999(35)  & 73   \\ 
\hline
$\beta=5.35$ & & & & & & & & &  & \\
\hline
   0.1270 & 16 &  0.0581(4) &  0.4124(66) &  0.429(22)  &  1.557(10) &  0.011 & 0.510 & 0.073 & 0.989(10) &  1.021(30)  & 30   \\ 
   0.1275 & 16 &  0.0396(4) &  0.3361(41) &  0.368(16)  &  1.729(10) &  0.268 & 0.112 & 0.435 & 0.986(7) &  1.046(27)  & 50   \\ 
   0.12775 & 16 &  0.0322(3) &  0.3115(49) &  0.376(12)  &  1.742(7) &  0.110 & 0.452 & 0.179 & 0.983(8) &  1.041(26)  & 110   \\ 
   0.1280 & 24 &  0.0237(4) &  0.2556(37) &  0.3485(60)  &  1.831(8) &  0.671 & 0.388 & -0.079 & 0.969(6) &  1.037(25)  & 150   \\ 
   0.1282 & 24 &  0.0169(2) &  0.2224(17) &  0.326(14)  &  1.888(5) &  0.443 & 0.351 & 0.115 & 0.962(6) &  1.059(23)  & 90   \\ 
   0.1283 & 24 &  0.0132(2) &  0.1916(28) &  0.3216(73)  &  1.937(8) &  0.339 & 0.396 & -0.096 & 0.962(6) &  1.085(20)  & 89   \\ 
   0.1284 & 24 &  0.0097(2) &  0.1650(66) &  0.2988(68)  &  1.982(6) &  -0.116 & 0.410 & -0.243 & 0.956(7) &  1.072(24)  & 90   \\ 
\hline
$\beta=5.4$ & & & & & & & & &  & \\
\hline
   0.1265 & 24 &  0.0565(9) &  0.3889(16) &  0.399(13)  &  2.028(9) &  -0.046 & 0.081 & -0.194 & 0.989(9) &  1.032(27)  & 90   \\ 
   0.1270 & 24 &  0.0410(2) &  0.3282(14) &  0.3631(77)  &  2.155(7) &  0.291 & 0.161 & -0.271 & 0.984(8) &  1.055(22)  & 90   \\ 
   0.1272 & 24 &  0.0339(2) &  0.2976(25) &  0.3463(35)  &  2.240(8) &  0.227 & 0.217 & 0.299 & 0.974(7) &  1.058(25)  & 90   \\ 
   0.1276 & 24 &  0.0203(6) &  0.2360(59) &  0.3050(30)  &  2.372(8) &  -0.157 & 0.002 & 0.015 & 0.973(6) &  1.076(20)  & 90   \\ 
   0.1277 & 24 &  0.0169(2) &  0.2102(20) &  0.2977(77)  &  2.430(10) &  0.188 & 0.489 & -0.208 & 0.971(8) &  1.046(26)  & 90   \\ 
   0.1278 & 24 &  0.0138(3) &  0.1855(21) &  0.2836(58)  &  2.467(10) &  0.402 & 0.032 & -0.075 & 0.967(6) &  1.063(24)  & 90   \\ 
   0.1279 & 24 &  0.0104(2) &  0.1631(58) &  0.2754(91)  &  2.574(13) &  0.218 & 0.696 & 0.172 & 0.970(6) &  1.112(18)  & 90   \\ 
\hline
 \end{tabular}
\caption{ Lattice data for $N_c=3$. The entries $C_1$, $C_2$ and $C_3$ correspond to the correlation coefficients
$C_{m_q,m_{PS}}$, $C_{m_q,f_{PS}}$, and $C_{m_{PS},f_{PS}}$.
\label{tab:data3}}
\end{table}

\begin{table}
\begin{tabular}{c c c c c c c c c c c c}
\hline
  $\kappa$ & $L$ & $am_q$  & $am_{PS}$ & $af_{PS}$ & $t_0/a^2$ &  $C_1$ & $C_2$ & $C_3$ & $Z_A$ & $Z_m$ & $N_{conf}$ \\
\hline
$\beta=10.0$ & & & & & & & & &  & \\
\hline
   0.1270 & 16 &  0.1017(4) &  0.6159(15) &  0.7086(77)  &  0.855(2) &  0.603 & 0.419 & 0.261 & 0.992(15) &  0.994(44)  & 61   \\ 
   0.1280 & 16 &  0.0570(10) &  0.4461(45) &  0.610(21)  &  1.023(4) &  0.620 & 0.187 & 0.258 & 0.984(14) &  1.011(41)  & 40   \\ 
   0.1285 & 16 &  0.0359(4) &  0.3456(22) &  0.5262(87)  &  1.131(5) &  0.522 & 0.018 & -0.207 & 0.974(10) &  1.038(36)  & 50   \\ 
   0.1288 & 16 &  0.0240(6) &  0.2798(44) &  0.483(14)  &  1.194(6) &  0.472 & 0.528 & 0.357 & 0.973(11) &  1.072(31)  & 90   \\ 
   0.1289 & 16 &  0.0192(4) &  0.2504(34) &  0.482(12)  &  1.230(7) &  0.421 & 0.636 & 0.284 & 0.974(12) &  1.053(33)  & 90   \\ 
   0.1290 & 16 &  0.0158(3) &  0.2264(29) &  0.489(14)  &  1.241(6) &  0.396 & 0.490 & -0.130 & 0.963(6) &  1.102(25)  & 80   \\ 
\hline
$\beta=10.1$ & & & & & & & & &  & \\
\hline
   0.1250 & 16 &  0.1161(1) &  0.6168(15) &  0.6138(96)  &  1.321(6) &  0.275 & 0.329 & 0.180 & 1.020(13) &  0.978(38)  & 90   \\ 
   0.1266 & 16 &  0.0623(2) &  0.4324(16) &  0.5255(71)  &  1.565(7) &  0.547 & 0.200 & 0.188 & 0.998(10) &  1.025(28)  & 80   \\ 
   0.1270 & 16 &  0.0481(2) &  0.3743(35) &  0.4905(52)  &  1.665(8) &  0.273 & 0.214 & 0.321 & 0.999(8) &  1.033(26)  & 80   \\ 
   0.1275 & 16 &  0.0311(2) &  0.2988(18) &  0.4391(97)  &  1.772(11) &  0.361 & -0.001 & 0.306 & 0.987(12) &  1.050(27)  & 90   \\ 
   0.12765 & 16 &  0.0264(1) &  0.2740(14) &  0.4420(41)  &  1.786(7) &  0.381 & 0.227 & -0.028 & 0.975(7) &  1.056(24)  & 140   \\ 
   0.12775 & 16 &  0.0226(2) &  0.2556(15) &  0.4132(76)  &  1.831(9) &  0.387 & -0.140 & -0.182 & 0.988(10) &  1.038(31)  & 120   \\ 
   0.1280 & 16 &  0.0147(5) &  0.2017(56) &  0.3955(78)  &  1.883(13) &  0.757 & -0.035 & -0.185 & 0.977(8) &  1.083(22)  & 50   \\ 
\hline
$\beta=10.2$ & & & & & & & & &  & \\
\hline
   0.1252 & 16 &  0.0866(2) &  0.4897(22) &  0.4957(28)  &  2.080(3) &  0.645 & 0.333 & 0.418 & 1.012(9) &  1.005(30)  & 190   \\ 
   0.1262 & 16 &  0.0547(1) &  0.3783(15) &  0.4473(35)  &  2.270(4) &  0.185 & 0.356 & 0.328 & 1.002(8) &  1.036(23)  & 190   \\ 
   0.1265 & 16 &  0.0454(2) &  0.3415(16) &  0.4314(50)  &  2.312(10) &  0.088 & 0.315 & -0.158 & 0.992(6) &  1.045(22)  & 90   \\ 
   0.1270 & 16 &  0.0290(2) &  0.2709(19) &  0.3881(44)  &  2.452(5) &  0.136 & 0.188 & -0.045 & 0.982(5) &  1.079(20)  & 101   \\ 
   0.1272 & 16 &  0.0222(1) &  0.2387(36) &  0.3659(97)  &  2.520(20) &  0.041 & 0.210 & -0.186 & 0.983(7) &  1.079(20)  & 90   \\ 
   0.1273 & 16 &  0.0191(2) &  0.2229(36) &  0.3555(94)  &  2.543(17) &  0.293 & 0.281 & -0.288 & 0.983(8) &  1.085(18)  & 90   \\ 
   0.1275 & 24 &  0.0124(1) &  0.1760(12) &  0.3429(41)  &  2.621(9) &  0.295 & 0.401 & -0.241 & 0.980(7) &  1.090(18)  & 90   \\ 
\hline
$\beta=10.3$ & & & & & & & & &  & \\
\hline
   0.1260 & 16 &  0.0494(3) &  0.3391(40) &  0.3910(62)  &  3.106(17) &  0.689 & 0.489 & 0.494 & 0.997(8) &  1.046(22)  & 50   \\ 
   0.1265 & 16 &  0.0337(1) &  0.2826(23) &  0.3503(53)  &  3.139(14) &  0.160 & 0.129 & -0.141 & 0.990(7) &  1.080(17)  & 98   \\ 
   0.12675 & 16 &  0.0255(1) &  0.2454(23) &  0.3347(49)  &  3.277(28) &  0.208 & 0.420 & -0.110 & 0.989(6) &  1.067(16)  & 90   \\ 
   0.1270 & 16 &  0.0174(1) &  0.2020(35) &  0.3124(85)  &  3.375(21) &  0.271 & -0.057 & -0.212 & 0.982(6) &  1.084(10)  & 90   \\ 
   0.1271 & 24 &  0.0142(1) &  0.1791(21) &  0.3044(34)  &  3.436(13) &  0.408 & -0.072 & -0.464 & 0.979(4) &  1.094(16)  & 90   \\ 
   0.1272 & 24 &  0.0108(1) &  0.1595(26) &  0.2918(35)  &  3.514(14) &  -0.282 & 0.148 & -0.015 & 0.978(7) &  1.079(20)  & 90   \\ 
\hline
 \end{tabular}
\caption{ Lattice data for $N_c=4$.  The entries $C_1$, $C_2$ and $C_3$ correspond to the correlation coefficients
$C_{m_q,m_{PS}}$, $C_{m_q,f_{PS}}$, and $C_{m_{PS},f_{PS}}$.
\label{tab:data4}}
\end{table}

\begin{table}
\begin{tabular}{c c c c c c c c c c c c}
\hline
 $\kappa$ & $L$ & $am_q$  & $am_{PS}$ & $af_{PS}$ & $t_0/a^2$ &  $C_1$ & $C_2$ & $C_3$ & $Z_A$ & $Z_m$ & $N_{conf}$ \\
\hline
$\beta=16.2$ & & & & & & & & &  & \\
\hline
   0.1250 & 16 &  0.1218(4) &  0.6461(23) &  0.7055(92)  &  1.161(4) &  0.421 & 0.448 & 0.507 & 1.029(16) &  0.950(44)  & 60   \\ 
   0.1260 & 16 &  0.0871(4) &  0.5309(20) &  0.6650(61)  &  1.290(7) &  0.545 & 0.426 & 0.083 & 1.020(14) &  0.988(38)  & 30   \\ 
   0.1270 & 16 &  0.0517(4) &  0.4011(22) &  0.599(15)  &  1.451(21) &  0.247 & -0.000 & 0.572 & 0.997(12) &  1.033(30)  & 30   \\ 
   0.1278 & 16 &  0.0249(2) &  0.2708(19) &  0.5250(65)  &  1.562(10) &  0.427 & 0.502 & 0.018 & 0.987(9) &  1.058(28)  & 70   \\ 
\hline
$\beta=16.3$ & & & & & & & & &  & \\
\hline
   0.1250 & 16 &  0.0982(3) &  0.5383(27) &  0.612(18)  &  1.709(7) &  0.414 & -0.087 & -0.027 & 1.033(12) &  0.979(33)  & 50   \\ 
   0.1260 & 16 &  0.0667(2) &  0.4377(17) &  0.5700(73)  &  1.823(6) &  -0.205 & -0.035 & 0.243 & 1.007(11) &  1.015(28)  & 50   \\ 
   0.1264 & 16 &  0.0543(1) &  0.3880(13) &  0.554(18)  &  1.871(7) &  0.375 & 0.227 & -0.112 & 1.010(11) &  1.031(25)  & 90   \\ 
   0.1268 & 16 &  0.0409(3) &  0.3335(12) &  0.514(14)  &  1.940(11) &  0.118 & 0.615 & 0.071 & 0.995(11) &  1.045(23)  & 90   \\ 
   0.1270 & 16 &  0.0346(1) &  0.3070(17) &  0.4992(51)  &  1.950(8) &  -0.033 & 0.336 & -0.199 & 0.996(9) &  1.056(24)  & 90   \\ 
   0.1273 & 16 &  0.0249(1) &  0.2551(12) &  0.4795(49)  &  2.030(8) &  0.079 & 0.243 & -0.065 & 0.996(8) &  1.060(22)  & 110   \\ 
   0.1275 & 16 &  0.0183(2) &  0.2220(23) &  0.4444(68)  &  2.057(7) &  0.299 & 0.141 & -0.004 & 0.986(7) &  1.066(21)  & 50   \\ 
\hline
$\beta=16.4$ & & & & & & & & &  & \\
\hline
   0.1252 & 16 &  0.0818(1) &  0.4701(7) &  0.5510(25)  &  2.181(4) &  0.167 & 0.183 & 0.160 & 1.015(11) &  1.006(28)  & 190   \\ 
   0.1258 & 16 &  0.0631(1) &  0.4037(14) &  0.5194(42)  &  2.272(5) &  -0.161 & 0.104 & 0.023 & 1.008(10) &  1.019(24)  & 190   \\ 
   0.1265 & 16 &  0.0409(1) &  0.3228(11) &  0.4792(55)  &  2.386(6) &  0.206 & 0.351 & 0.242 & 1.000(11) &  1.047(23)  & 190   \\ 
   0.1270 & 16 &  0.0247(1) &  0.2478(16) &  0.4283(25)  &  2.483(6) &  0.161 & 0.114 & -0.146 & 0.986(9) &  1.081(18)  & 200   \\ 
   0.1272 & 24 &  0.0183(1) &  0.2122(8) &  0.4236(68)  &  2.526(6) &  0.070 & 0.074 & -0.128 & 0.987(10) &  1.082(19)  & 90   \\ 
\hline
$\beta=16.6$ & & & & & & & & &  & \\
\hline
   0.1252 & 16 &  0.0695(1) &  0.4040(12) &  0.4705(37)  &  3.136(12) &  0.255 & 0.292 & 0.063 & 1.020(11) &  1.024(24)  & 90   \\ 
   0.1260 & 16 &  0.0443(1) &  0.3186(15) &  0.4171(45)  &  3.360(22) &  0.164 & 0.102 & 0.197 & 1.005(7) &  1.051(21)  & 110   \\ 
   0.1264 & 16 &  0.0318(1) &  0.2674(15) &  0.3934(37)  &  3.407(22) &  0.253 & 0.037 & 0.029 & 0.993(4) &  1.067(18)  & 140   \\ 
   0.1266 & 16 &  0.0254(1) &  0.2382(22) &  0.3712(63)  &  3.455(16) &  -0.243 & -0.133 & 0.060 & 1.002(9) &  1.078(16)  & 140   \\ 
   0.1269 & 24 &  0.0159(0) &  0.1814(12) &  0.3591(31)  &  3.543(12) &  0.281 & 0.257 & -0.282 & 0.988(7) &  1.093(16)  & 90   \\ 

\hline
 \end{tabular}
\caption{ Lattice data for $N_c=5$.  The entries $C_1$, $C_2$ and $C_3$ correspond to the correlation coefficients
$C_{m_q,m_{PS}}$, $C_{m_q,f_{PS}}$, and $C_{m_{PS},f_{PS}}$.
\label{tab:data5}}
\end{table}

\begin{table}
\begin{tabular}{c c }
\hline
$\beta$ & $t_0/a^2$ \\
\hline
$SU(3)$ &  \\
\hline
5.2 &  1.037(7) \\
5.3 &  1.374(13) \\
5.3 &  1.804(17) \\
5.4 &  2.379(19) \\
5.5 &  3.033(30) \\
\hline
$SU(4)$ &  \\
\hline
10.0 &  1.116(14) \\
10.1 &  1.770(12) \\
10.2 &  2.503(33) \\
10.3 &  3.339(25) \\
\hline
$SU(5)$ &  \\
\hline
16.2 &  1.510(23) \\
16.3 &  1.959(10) \\
16.4 &  2.468(8) \\
16.6 &  3.487(27) \\
\hline
\end{tabular}
\caption{Flow parameter values at $\xi=0.12$.
\label{tab:t0}}
\end{table}

\begin{table}
\begin{tabular}{c c c c}
\hline
  & $N_c=3$ &  $N_c=4$ & $N_c=5$ \\
\hline
$\sqrt{t_0}B$ & 1.90(10) &   1.98(8) &  1.82(7) \\
$\sqrt{t_0}F$ & 0.086(5) &   0.111(5) &  0.137(4) \\
$l_4$ & 4.2(7) &   4.9(6) &  5.2(4) \\
$l_3$ & $3.2\pm 1.5$ &   $3.62\pm 1.3$ & $2.9\pm 1.3$ \\
\hline
$c_B$ & -0.14(3) &   -0.20(2) &  -0.14(4) \\
$c_F$ & -0.6(2) &   -0.06(2) &  -0.04(3) \\
\hline
$l_{12}$ & 2.7(5) &   3.1(4) &  3.3(5) \\
$k_F$ & $1.14 \pm 1.5$  & $1.0 \pm 1.4$  &  $1.5\pm 1.8$  \\
$k_M$ & $0.15\pm 1.0$  & $0.4\pm 0..8$  & $0.5\pm 1.0$  \\
\hline
 \end{tabular}
\caption{ LO and NLO LEC's  of  $SU(2)$ chiral perturbation theory from our $N_c=3$, 4, and 5 data sets.
\label{tab:xlec}}
\end{table}

\begin{table}
\begin{tabular}{c c c c  c}
\hline
  &  $N_c=4$ & converted  & $N_c=5$ & converted \\
\hline
$\sqrt{t_0}B$ & 1.84(9) & 1.96(10) &  1.65(6) & 1.79(7) \\
$\sqrt{t_0}F$ & 0.109(4) &  &  0.136(4) & \\
$l_4^{0}+l_4^{(1)}/N_c$ & 0.22(6) & 4.93(24)  &  0.28(6) & 5.45(30) \\
$l_3^{(0)}+l_3^{1}/N_c$ & -0.74(24) & $-0.57\pm 1.15$  &  -0.81(23) & $-0.11\pm 0.23$ \\
\hline
$c_B$ & -0.20(2) &  &  -0.14(4) & \\
$c_F$ & -0.06(2) &  &  -0.04(3) & \\
\hline
$T_F$ & 0.08(5) &  &  0.06(4) & \\
$T_M$ & -0.06(7) &  &  -0.07(6) & \\
\hline
 \end{tabular}
\caption{ LO and NLO LEC's  of  $U(2)$ chiral perturbation theory from our $N_c=4$  and 5 data sets.
The ``converted'' columns show conversions to $SU(2)$ quantities according to Eq.~\ref{eq:unsus}.
In the column we have listed $l_3$ and $l_4$; note the rescaling.
\label{tab:ulec}}
\end{table}
\clearpage

\begin{acknowledgments}
We would like to thank 
William Jay
and
Fernando Romero-L\'opez
 for correspondence.
Our computer code is based on the publicly available package of the                                   
 MILC collaboration~\cite{MILC}. The version we use was originally developed by Y.~Shamir and         
 B.~Svetitsky.                                                                                       
This material is based upon work supported by the U.S. Department of Energy, Office of Science, Office of
High Energy Physics under Award Number DE-SC-0010005.                                                    
Some of the computations for this work were also carried out with resources provided by the USQCD        
Collaboration, which is funded                                                                           
by the Office of Science of the U.S.\ Department of Energy                                               
using the resources of the Fermi National Accelerator Laboratory (Fermilab), a U.S.                      
Department of Energy, Office of Science, HEP User Facility. Fermilab is managed by                       
 Fermi Research Alliance, LLC (FRA), acting under Contract No. DE- AC02-07CH11359.                       
\end{acknowledgments}



\end{document}